\documentclass[twocolumn]{aastex631}

\usepackage{longtable}

\begin{document}

\title{Statistical Analysis of PAHs as a Tracer of Anomalous Microwave Emission Using DIRBE Data}

\correspondingauthor{Danielle Sponseller}

\author[0000-0003-3922-1487]{Danielle Sponseller}
\affiliation{Department of Physics \& Astronomy, Johns Hopkins University, Baltimore, MD 21218, USA}
\affiliation{Department of Space, Earth and Environment, Chalmers University of Technology, Gothenburg, Sweden}
\email{danielle.sponseller@chalmers.se}

\author[0000-0003-0016-0533]{David T. Chuss}
\affiliation{Department of Physics, Villanova University, Villanova, PA 19085, USA}

\author[0000-0001-7449-4638]{Brandon Hensley}
\affiliation{Jet Propulsion Laboratory, California Institute of Technology, 4800 Oak Grove Drive, Pasadena, CA 91109, USA}

\author[0000-0001-9835-2351]{Alan Kogut}
\affiliation{Code 665, Goddard Space Flight Center, Greenbelt, MD 20771, USA}

\begin{abstract}

We use archival data from the Diffuse Infrared Background Experiment (DIRBE) to map the polycyclic aromatic hydrocarbon (PAH) 3.3 $\mu$m emission feature and analyze its correlation with AME in 98 compact sources identified by the \textit{Planck} collaboration.
We find that while FIR thermal dust emission continues to be a better tracer of AME in most of the considered regions, 17\% of the AME sources are better correlated with emission from small PAHs as traced by DIRBE.
Furthermore, of the 27 sources which were identified as highly significant AME detections in the \textit{Planck} analysis, 37\% prefer PAHs as an AME tracer.
Further work is required to understand to what extent local interstellar conditions are affecting PAH emission mechanisms and to reveal the underlying carriers of AME.

\end{abstract}

\keywords{Interstellar Medium (847) --- Polycyclic Aromatic Hydrocarbons (1280) --- Dust Physics (2229)}

\section{Introduction}

Cosmic microwave background (CMB) experiments aimed at testing inflation must precisely characterize and remove astrophysical foregrounds.
When the known foregrounds of thermal dust, synchrotron, and free-free emission are subtracted, a residual signal remains which is referred to as anomalous microwave emission (AME) \citep{Kogut1996a, Leitch1997}.
Despite having been discovered over 25 years ago, AME has not yet been uniquely identified as originating from any known constituent of the ISM \citep{Dickinson2018}.
This excess emission is known to be spatially correlated with thermal dust emission, however its spectrum does not match the expected dust emission at low frequencies.
Currently, the best candidate for AME is electric dipole radiation from ultra-small rapidly rotating dust grains \citep{Draine1998a}.

Though constrained to be less than 1\% polarized \citep{Dickinson2011, Genova-Santos2017, Herman2023}, AME may still be a relevant foreground to the increasingly sensitive next-generation CMB telescopes searching for primordial B-modes \citep{Remazeilles2016}.
Even if this foreground is entirely unpolarized, its characterization is vital to allow future experiments targeting CMB spectral distortions to separate it from synchrotron emission \citep{Abitbol2017a}.
Understanding the physics of AME is thus vital for improving foreground models.

\cite{Dickinson2018} provide a review of AME research and describe several potential physical mechanisms for the origin of this emission.
These include electric dipole radiation from spinning dust grains \citep{Draine1998a}, thermal emission from dust grains with magnetic inclusions \citep{Draine1999, Draine2013}, variations in the optical properties of amorphous materials, a two level system (TLS) model for low-temperature amorphous grains \citep{Meny2007, Jones2009, Nashimoto2020}, or optically thick free-free emission from warm ionized gas.
However, the most plausible of these candidates remains the spinning dust hypothesis.

If spinning dust is indeed responsible for AME, the current best candidate carrier is a species of grains known as polycyclic aromatic hydrocarbons (PAHs).
PAHs have the right size distribution \citep{Li2001} and can acquire the necessary electric dipole moments (e.g. through carbon substitutions) to account for the observed emission.
They are also known to be ubiquitous throughout the interstellar medium (ISM) \citep{Tielens2008}.

Several authors have investigated spatial correlations between the \textit{Planck} AME map and potential tracers.
\cite{Hensley2016} examine PAH emission as traced by the Wide-field Infrared Survey Explorer (WISE) W3 band, which is centered around 12 $\mu$m and encompasses the 7.7, 8.6, 11.3, 12, 12.7, 13.55, and 17 $\mu$m features, and compare the AME correlation with this PAH emission to its correlation with far-IR thermal dust emission over large parts of the sky.
They find that while PAHs are well-correlated with AME, thermal dust emission is still a better tracer, contrary to the expectations of the PAH hypothesis.
\cite{Chuss2022} use data from the Diffuse Infrared Background Experiment (DIRBE) mission to map PAH emission at 3.3 $\mu$m within a single prominent AME region, $\lambda$-Orionis.
Similarly, they conclude that AME is better spatially traced by thermal dust emission than by PAHs.
However, these authors emphasize that it is possible that variations in interstellar environments may be affecting PAH emission mechanisms, which may explain the observed low AME-PAH correlation, even if PAHs are the cause of AME.

On the other hand, there is some evidence in favor of PAHs over thermal dust emission.
\cite{Bell2019} use data from the AKARI/Infrared Camera to study the $\lambda$-Orionis region and find that AME is better correlated with PAH emission at 9 $\mu$m than with dust mass.
They also find that PAH mass is more highly correlated with AME than is total dust mass.
\cite{Planck-Collaboration2011} consider the Perseus Molecular Cloud and $\rho$-Ophiuchus regions and find evidence that AME likely originates from a relatively dense gas component of the ISM, and could be due to PAHs.

\cite{Planck-Collaboration2014a} identify bright compact sources and obtain spectral energy distributions (SEDs) from aperture photometry to produce a list of 98 candidate AME sources.
The data used span a frequency range of 0.408 to 37,500 GHz based on \textit{Planck} as well as ancillary datasets.
The SEDs are used to perform a parametric fit with CMB, thermal dust emission, synchrotron, free-free, and AME components.
For this analysis, the AME spectrum is based on a one-component spinning dust model assuming warm ionized medium conditions, and the peak of the spectrum is allowed to vary via shifts in frequency.
This analysis finds evidence in favor of very small dust grains and PAHs as carriers of AME.

Many of the bright AME sources contain free-free emission, which is often observed in ultra-compact HII (UCHII) regions.
Free-free emission must be considered carefully when measuring AME as it can be difficult to separate these components.
A subset of 27 AME sources were identified as highly statistically significant based on a signal-to-noise ratio (SNR) of at least 5 as well as a maximal fractional contribution of UCHII regions to the total flux density of 0.25.
Of the highly significant AME regions, most lie outside the Galactic plane.

\cite{Poidevin2023} use aperture photometry to study a subset of 42 of the \textit{Planck} AME sources and 10 additional molecular cloud regions, finding a strong correlation between the peak flux densities of the AME and thermal dust emission.
Note that while the \cite{Planck-Collaboration2014a} and \cite{Poidevin2023} analyses make statistical comparisons across a large sample of sources, they do not analyze the spatial correlations between AME and other tracers within individual regions.
Of the studies that do examine spatial correlations, most only focus on one or a small number of sources.

In this paper, we use the set of 98 candidate AME sources identified by \cite{Planck-Collaboration2014a} to examine the relationship between AME and PAH emission in a large sample of sources across different astrophysical environments.
We use DIRBE data to produce maps of PAH emission at 3.3 $\mu$m within each of the candidate AME sources.
We then perform a spatial correlation analysis between AME, PAHs, and thermal dust emission within a 4 degree by 4 degree square patch centered on each of the 98 regions, significantly increasing the number of sources for which these correlations have been quantified.
Note that while these regions were originally identified based on a fit to a single-component AME model, we use the spectral template and fitted parameters of the two-component AME model provided in the \textit{Planck} Commander map to evaluate the total AME intensity at 30 GHz.
We additionally perform a correlation analysis on the well-studied $\lambda$-Orionis region, however in this case we use a 17 degree by 17 degree square patch as this is a much more extended source.

Our aim with this spatial correlation analysis is to test the hypothesis that PAHs are the carrier of AME.
In section \ref{sec:data} we discuss the data used.
In section \ref{sec:PAH_emission}, we describe our procedure for generating maps of the PAH 3.3 $\mu$m feature within each of the considered regions.
We then analyze the AME-PAH correlations in section \ref{sec:results}.
Finally, we summarize our findings in section \ref{sec:conclusions}.

\section{Data}
\label{sec:data}

To assess the correlations between AME and other ISM constituents, we require spatial maps of AME, thermal dust emission, and PAH emission in each region.
For the AME, we use published \textit{Planck} AME maps and evaluate the spectral template to obtain a map of the total AME intensity at 30 GHz.
For thermal dust emission, we use the \textit{Planck} 857 GHz map.
This map is downgraded to a resolution of $N_{\mathrm{side}}=256$ to match the resolution of the DIRBE maps and the \textit{Planck} AME map for the correlation analysis.

The peak frequency of the AME spectrum generally occurs between 10--60 GHz.
\cite{Planck-Collaboration2016d} perform a component separation analysis by performing spectral fitting with parametric models to create full-sky maps of Galactic foregrounds, including AME, using the Commander code \citep{Eriksen2008}.
The fitted AME spectrum is based on a two-component spinning dust model where one component has a fixed peak frequency while the other allows for a variable peak frequency via rigid translations in log-log space.
This effectively parameterizes the width of the AME spectrum.
The spectral template used for both AME components is based on the predicted spinning dust emission in cold neutral medium (CNM) conditions, with a hydrogen density of $n_{\mathrm{H}} = 30$ cm$^{-3}$, a temperature of $T = 100$ K, a hydrogen ionization fraction of $n_{\mathrm{H}^{+}}/n_{\mathrm{H}} = 10^{-3}$, a relative abundance of ionized carbon given by $n_{\mathrm{C}^{+}}/n_{\mathrm{H}} = 3 \times 10^{-4}$, no H$_2$ formation ($\gamma = 0$), and a radiation field strength (defined relative to the ISM average) of $\chi \equiv u/u_{\mathrm{ISRF}} = 1$ \citep{Ali-Haimoud2009}.
The average ISRF spectrum is estimated by \cite{Mezger1982} and \cite{Mathis1983}, resulting in a total energy density of $8.64 \times 10^{-13}$ erg cm$^{-3}$ for $h\nu < 13.6$ eV \citep{Weingartner2001}.
While this two-component AME model has been successfully applied in a variety of astrophysical environments, it is not based on an understanding of the underlying physics of AME.

To obtain a map of AME at 30 GHz, we use the best-fit parameters and amplitudes for the two AME components along with the accompanying spectral templates described in \cite{Planck-Collaboration2016d} to obtain the total combined AME at 30 GHz.
We note that one potential limitation in the AME map is its separation from the synchrotron component, as spectral index variations complicate the precise fitting of the synchrotron emission \citep{Planck-Collaboration2016e}.

\begin{figure*}
    \centering
    \includegraphics[width=\textwidth]{"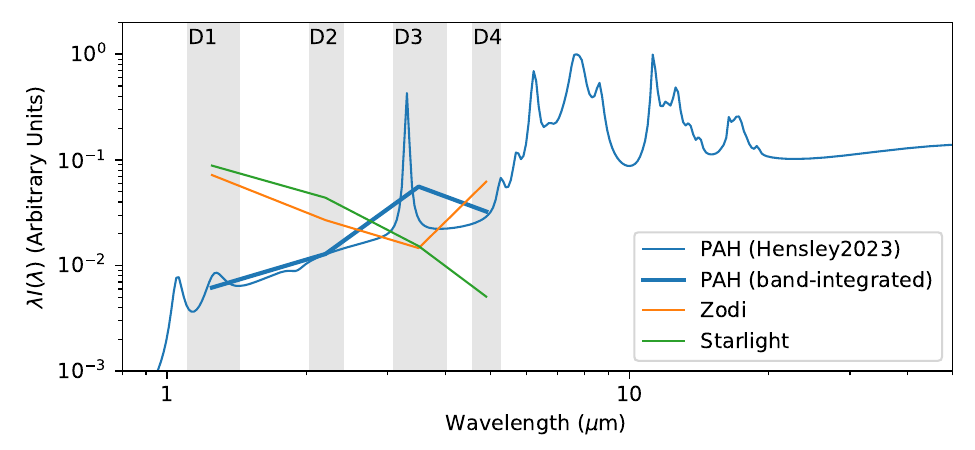"}
    \caption{A representative PAH emission spectrum is shown as a thin, blue line. DIRBE Bands 1--4 are shown in gray. Band 3 contains the prominent 3.3 $\mu$m emission feature while the adjacent bands contain only continuum emission from PAHs, allowing this feature to be isolated. The band-integrated PAH spectrum used to perform the linear least-squares fitting is shown as a thicker blue line. The zodiacal emission and starlight spectral basis functions are shown as thin lines. These latter two components show an example for an arbitrary pixel, as these are calculated on a pixel-by-pixel basis using the IPD and FSM models, respectively.}
    \label{fig:PAH_spect}
\end{figure*}

To obtain a PAH map, we utilize data from the DIRBE instrument onboard the Cosmic Background Explorer (COBE) satellite to fit for PAH emission.
DIRBE obtained data in ten bands from 1.25 to 240 $\mu$m with a spatial resolution of 0.7 degrees.
Its first four bands (1.25, 2.2, 3.5, and 4.9 $\mu$m) are dominated by starlight and Zodiacal emission, with the third band also containing a prominent PAH feature, the 3.3 $\mu$m C--H stretching mode, as shown in Figure \ref{fig:PAH_spect}.
This allows us to expand on the method used in \cite{Chuss2022} by performing a full spectral fit of these three emission components in each pixel using DIRBE bands 1--4.
We do not include longer-wavelength bands as they are contaminated by additional foregrounds such as thermal dust emission.

We use the Zodi-Subtracted Mission Average (ZSMA) DIRBE sky maps\footnote{\url{http://cade.irap.omp.eu/dokuwiki/doku.php?id=dirbe}}.
Zodiacal emission is a combination of scattering as well as thermal emission from interplanetary dust within our solar system, and the observed contribution of this component varies significantly over the course of the DIRBE mission.
The ZSMA maps were produced by subtracting estimates of zodiacal emission from DIRBE intensity maps on a weekly basis and then averaging the remaining residual maps together \citep{Kelsall1998}.
While we use the zodi-subtracted maps to estimate PAH emission, we still fit a zodiacal component as some residual emission remains.

\section{Estimating PAH Emission}
\label{sec:PAH_emission}

Before performing our correlation analysis, we first mask out bright stellar point sources, as the starlight component in our fitted model accounts for diffuse emission due to the distribution of stars, but does not model individual point sources well.
To do this, we calculate the background level in DIRBE Band 1 based on the median value within an annulus of 0.75 -- 1.0 degrees centered around each pixel.
Note that the lower bound of the annulus size is larger than the DIRBE beam size of 0.7 degrees.
We then mask all pixels where the Band 1 intensity is at least 0.5 MJy sr$^{-1}$ greater than the local background level.

To obtain the PAH 3.3 $\mu$m emission in each pixel, we perform a spectral fit of the three components relevant to DIRBE's four lowest-wavelength bands -- starlight, zodiacal emission, and PAH emission.
Amplitudes for these components are fit in each pixel using the linear least-squares (LLS) technique.

Starlight is modeled using the Faint Source Model (FSM), a statistical model of unresolved stellar emission which was specifically designed for DIRBE analysis.
The model takes into account the distribution of stars in the Milky Way, reddening due to dust, as well as a distribution of the types and temperatures of stars.
A predicted spectrum is given in each pixel of the sky represented in the quad tree pixelization scheme.
For each HEALPix\footnote{\url{https://healpix.sourceforge.io}} pixel \citep{Gorski2005}, the FSM model prediction in the nearest skycube pixel was used.
The DIRBE maps have $N_{\mathrm{side}} = 256$, which corresponds to a pixel size of approximately 0.23 degrees.

The Zodiacal emission spectrum in each pixel was computed from the interplanetary dust (IPD) model \citep{Kelsall1998}.
For each skycube pixel, the average of all available weekly IPD model spectra predictions was computed.
The HEALPix map was then generated based on the nearest skycube pixel.
The effects of differing pixelization schemes are expected to be negligible.

The PAH emission spectrum shown in Figure \ref{fig:PAH_spect} was obtained from models\footnote{\href{https://dataverse.harvard.edu/dataverse/astrodust}{https://dataverse.harvard.edu/dataverse/astrodust}} provided in \cite{Hensley2023}.
This model is based on a fiducial grain size distribution at high Galactic latitudes that reproduces the average extinction curve within the Milky Way.
To obtain a PAH spectral basis function for our LLS fitting procedure, we integrate the SED over each of the DIRBE bands to evaluate the spectrum as it would be observed by DIRBE.
We show representative spectra for all three fitted components in Figure \ref{fig:PAH_spect}.

We now use the PAH, starlight, and Zodiacal spectral basis functions described above to map the PAH 3.3 $\mu$m emission in each region.
The linear least-squares method is used to fit amplitudes for all components within each pixel.
We use $X_1(\nu), \ldots, X_N(\nu)$ to represent our spectral basis functions, where in our case $N=3$.
Our goal is to construct a model $Y(\nu)$ from a linear combination of these basis functions
\begin{equation}
	Y(\nu_i) = \sum_{k = 1}^{N} a_k X_k(\nu_i)
\end{equation}
which best represents our observations within each pixel.
To do this, we first define the goodness-of-fit metric
\begin{equation}
	\chi^2 = \sum_{i = 1}^{M} \left[ \frac{y_i - Y(\nu_i)}{\sigma_i} \right]^2
\end{equation}
where $y_i$ is the observed intensity in DIRBE channel $\nu_i$ and $\sigma_i$ is the measurement noise.
Our goal then is to find the model parameters $a_k$ such that $\chi^2$ is minimized.
In the linear least-squares method, we do this by taking the partial derivatives of the $\chi^2$ function with respect to each parameter and setting them equal to zero, producing a set of $N$ equations.
We can define the design matrix $A$ as
\begin{equation}
	A_{ij} = \frac{X_j(\nu_i)}{\sigma_i}
\end{equation}
and a vector $\vec{b}$ as
\begin{equation}
	b_i = \frac{y_i}{\sigma_i}.
\end{equation}
We can solve for the parameters $\vec{a}$ by using the normal equations, which can be written as
\begin{equation}
	\left( A^T \cdot A \right) \cdot \vec{a} = A^T \cdot \vec{b}
\end{equation}
\citep{Press1992}.
Finally, we can define the covariance matrix $C$ as
\begin{equation}
	C = \left( A^T \cdot A \right)^{-1}
\end{equation}
where the square roots of the diagonal elements are the parameter uncertainties.
The starlight and PAH amplitudes are constrained to be non-negative, while the Zodiacal emission amplitude is unconstrained as we are using the Zodi-Subtracted Mission Average (ZSMA) maps and are only including this component to account for any over- or under-subtraction in a given pixel.
To implement these constraints, we use the bounded linear least-squares optimization function \texttt{lsq\_linear} provided by the \textsc{scipy} Python package.

\begin{figure*}
    \centering
    \includegraphics[width=\textwidth]{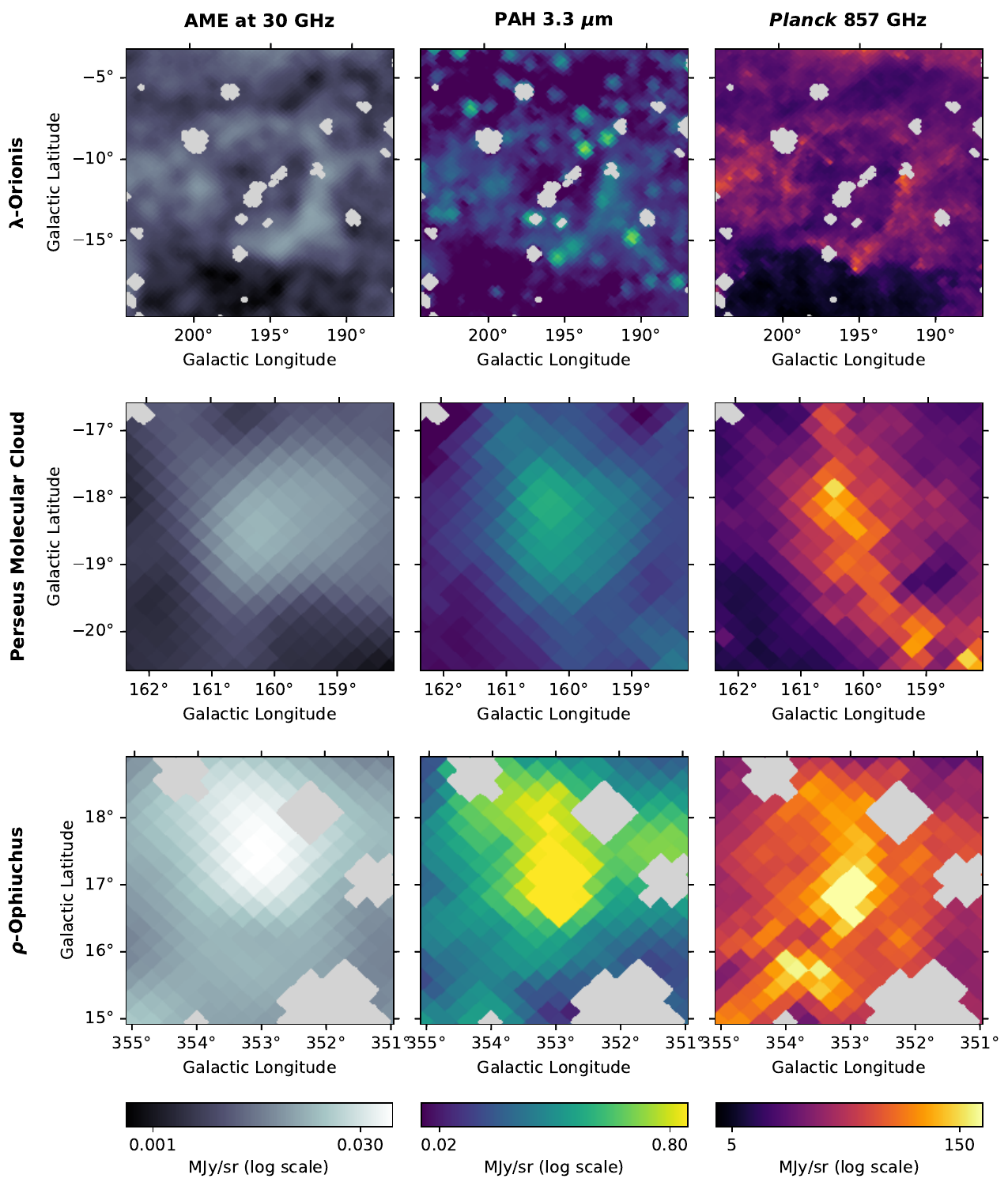}
    \caption{Maps of $\lambda$-Orionis, the Perseus Molecular Cloud, and $\rho$-Ophiuchus are shown in the top, middle, and bottom rows, respectively. The left column is the AME at 30 GHz, the middle shows our map of PAH 3.3 $\mu$m emission, and the right column shows thermal dust emission as traced by the \textit{Planck} 857 GHz map degraded to the resolution of the DIRBE maps ($N_{\mathrm{side}}=256$). Masked point sources are shown in gray.}
    \label{fig:reg_maps}
\end{figure*}

Having obtained a PAH map, we show maps of the total AME at 30 GHz, the PAH 3.3 $\mu$m emission, and the thermal dust emission at 857 GHz for a few select AME sources in Figure \ref{fig:reg_maps}.
Masked pixels are shown in gray.
We focus in particular on $\lambda$-Orionis \citep{Harper2025}, the Perseus Molecular Cloud \citep{Genova-Santos2015}, and $\rho$-Ophiuchus \citep{Arce-Tord2020}, all of which are well-studied ISM regions with AME.

The $\lambda$-Orionis region contains a bright ring-like structure with an angular diameter of roughly ten degrees centered on the star $\lambda$-Orionis.
\cite{Maddalena1987} identify the ring as being composed of expanding molecular gas and dust clouds.
It is also known to contain a prominent photodissociation region (PDR).
In the top row of Figure \ref{fig:reg_maps}, we see that this structure is visible in all three maps, suggesting good spatial correlation between all components.

Perseus, a large complex extending over more than 150 pc, contains a dense molecular cloud centered on $(l, b) = (160.26, -18.62)$ which is known to be one of the prominent AME sources within our Galaxy \citep{Tibbs2011}.
This region displays a distributed dust structure, as shown in the right panel of the middle row of Figure \ref{fig:reg_maps}.
On the other hand, it appears as a single bright source without much structure in either AME or PAH emission.

The $\rho$-Ophiuchus region features a translucent molecular cloud.
While there is a PDR, it does not appear to contain an HII region, unlike $\lambda$-Orionis \citep{Casassus2021}.
In maps of $\rho$-Ophiuchus in the bottom row of Figure \ref{fig:reg_maps}, a single bright source is observed in all components, however there are spatial offsets in the location of the peak emission.

\begin{figure}
    \centering
    \includegraphics[width=\columnwidth]{"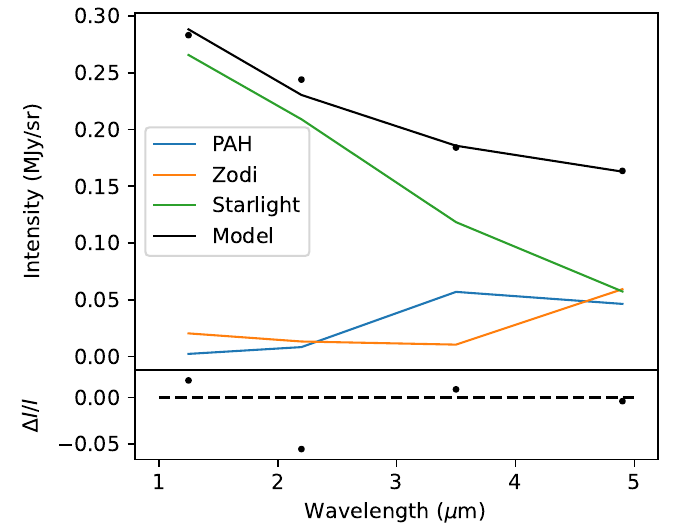"}
    \includegraphics[width=\columnwidth]{"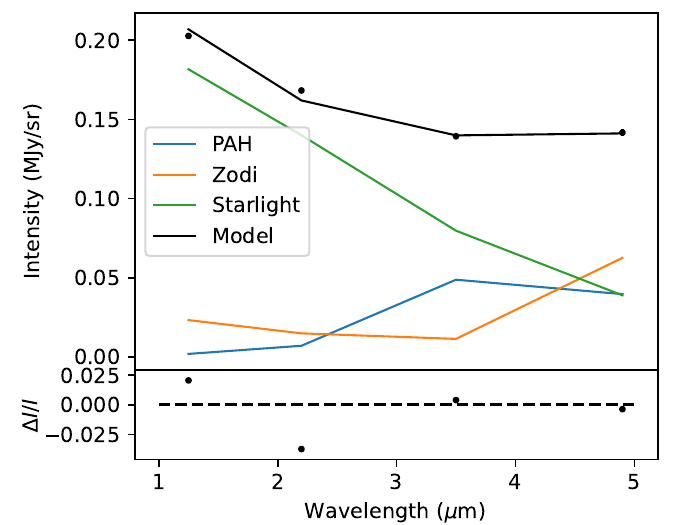"}
    \includegraphics[width=\columnwidth]{"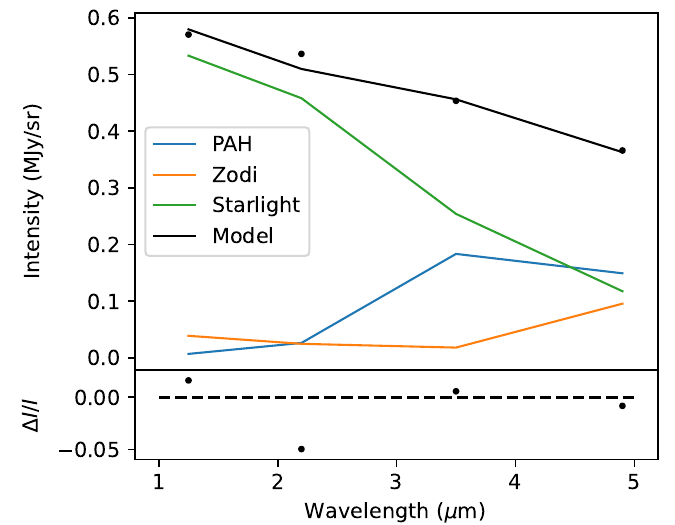"}
    \caption{Examples of spectral fits in $\lambda$-Orionis, the Perseus Molecular Cloud, and $\rho$-Ophiuchus are shown in the top, middle, and bottom panels, respectively. Each panel shows the fit for a randomly-chosen unmasked pixel. DIRBE data and error bars are shown as well, however the error bars are smaller than the marker points. The root-mean-square fractional residuals are on the order of 2--3\% in each case. The locations of the pixels in Galactic coordinates are $(l, b) = (202.2, -14.0)$, $(158.4, -18.8)$, and $(354.2, 16.3)$, respectively.}
    \label{fig:example_fits}
\end{figure}

We show example spectral fits chosen randomly from an unmasked pixel within each of these three regions in Figure \ref{fig:example_fits}.
The RMS fractional residuals are on the order of 2--3\% in each case.

\section{Analyzing the Correlation Between AME and PAH Emission}
\label{sec:results}

In this section we analyze the correlation between AME and the PAH 3.3 $\mu$m feature over the 98 \textit{Planck} AME sources. Pixels within a $4 \times 4$ degree square patch are used for each of the \textit{Planck} AME sources (approximately 300 pixels per region).
For $\lambda$-Orionis, we use a $17 \times 17$ degree square patch (approximately 5000 pixels).
We use the Spearman rank correlation coefficient $r$ to determine the correlation between AME and PAH emission, AME and dust emission, and PAH vs. thermal dust emission.
We use the Spearman correlation instead of the Pearson correlation since the Spearman test depends only on a monotonic relationship between variables and does not assume a linear scaling.
Pixels masked due to the presence of bright stellar sources in Band 1 are not included in the calculation of the correlation coefficients.

\begin{figure*}
    \centering
    \includegraphics[width=\textwidth]{"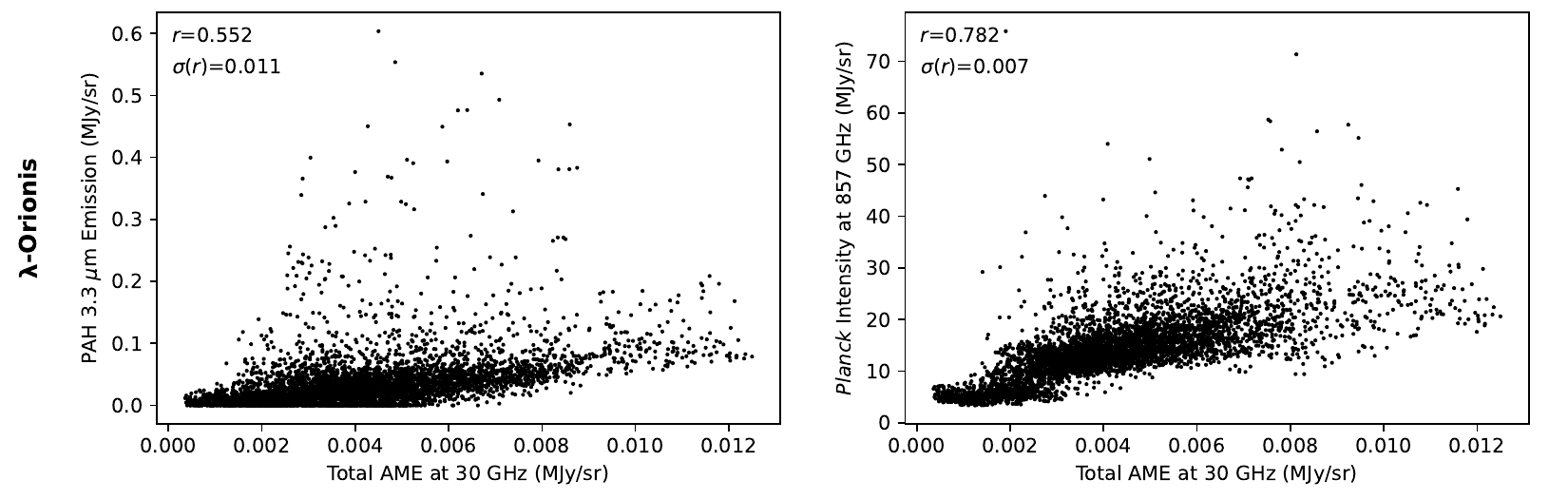"}
    \includegraphics[width=\textwidth]{"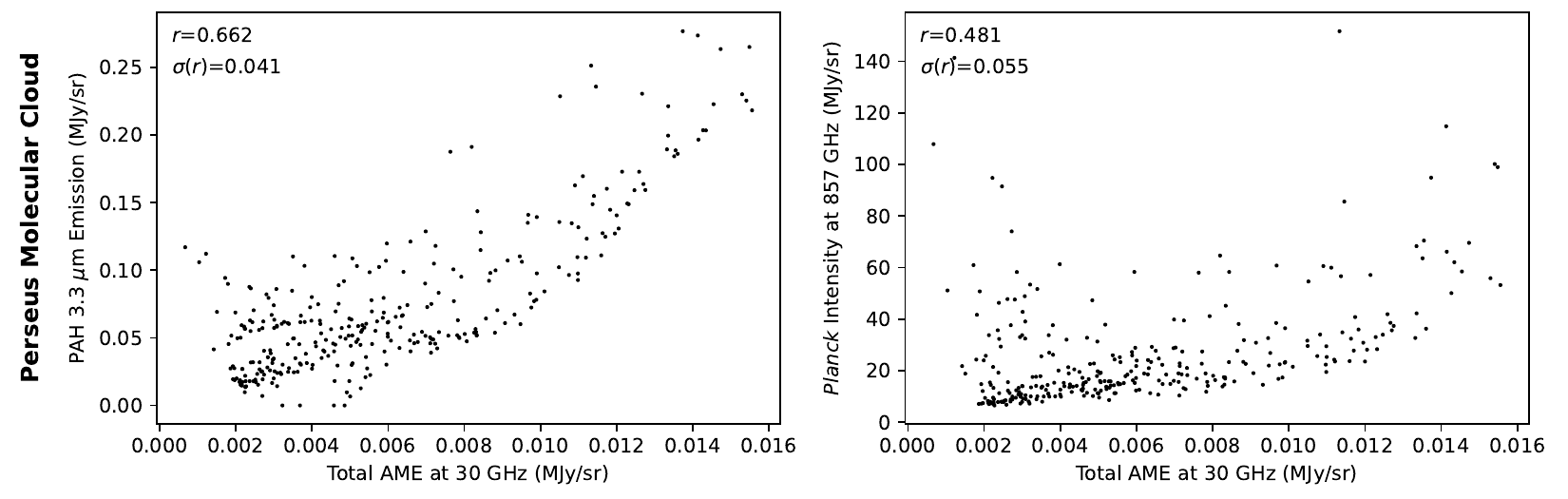"}
    \includegraphics[width=\textwidth]{"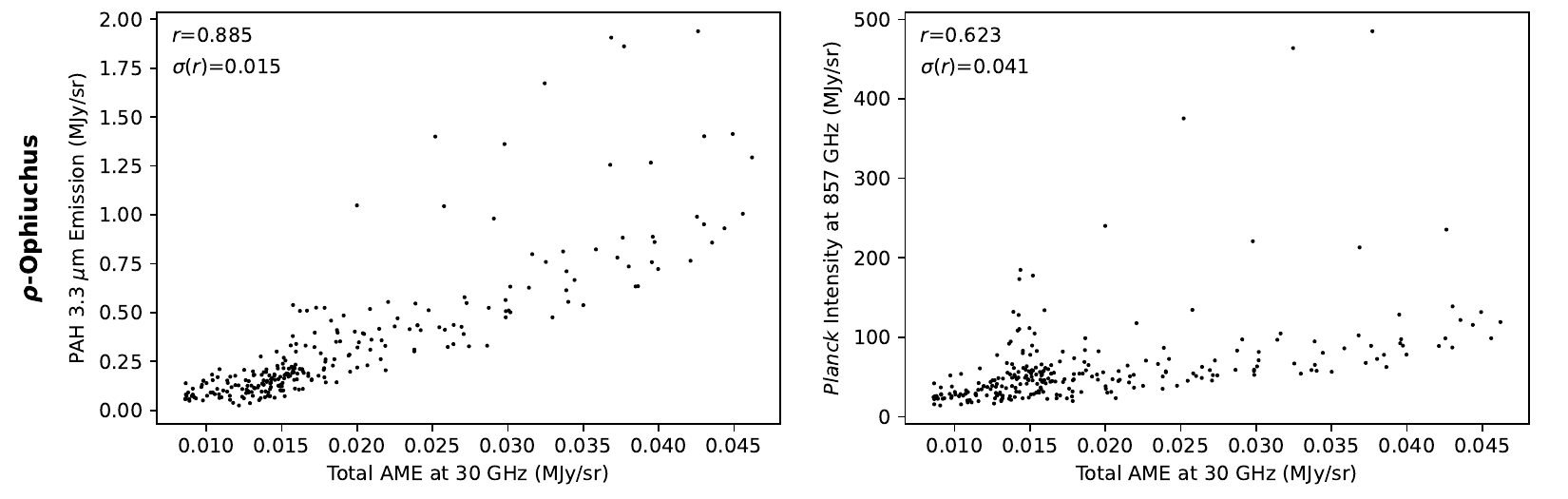"}
    \caption{Scatter plots for $\lambda$-Orionis, the Perseus Molecular Cloud, and $\rho$-Ophiuchus (top, middle, and bottom rows, respectively). The AME-PAH correlations are shown in the left column while the AME-dust correlations are shown on the right. The Spearman rank correlation coefficient is shown in the upper left of each panel. In $\lambda$-Orionis, we find that AME is better correlated with thermal emission from larger dust grains, while AME is better correlated with PAHs in the Perseus Molecular Cloud and $\rho$-Ophiuchus.}
    \label{fig:scatter}
\end{figure*}

Scatter plots showing the AME correlation with the PAH 3.3 $\mu$m feature and thermal dust emission for $\lambda$-Orionis, the Perseus Molecular Cloud, and $\rho$-Ophiuchus are shown in Figure \ref{fig:scatter}.
AME (``A'') in the $\lambda$-Orionis region is better correlated with thermal dust emission (``D'') than with PAHs ( ``P''; $r_{\mathrm{AP}}=0.552$, $r_{\mathrm{AD}}=0.782$) while PAHs appear to be the preferred tracer in the Perseus Molecular Cloud ($r_{\mathrm{AP}}=0.662$, $r_{\mathrm{AD}}=0.481$) and $r$-Ophiuchus ($r_{\mathrm{AP}}=0.885$, $r_{\mathrm{AD}}=0.623$).

\begin{figure}
    \centering
    \includegraphics[width=\columnwidth]{"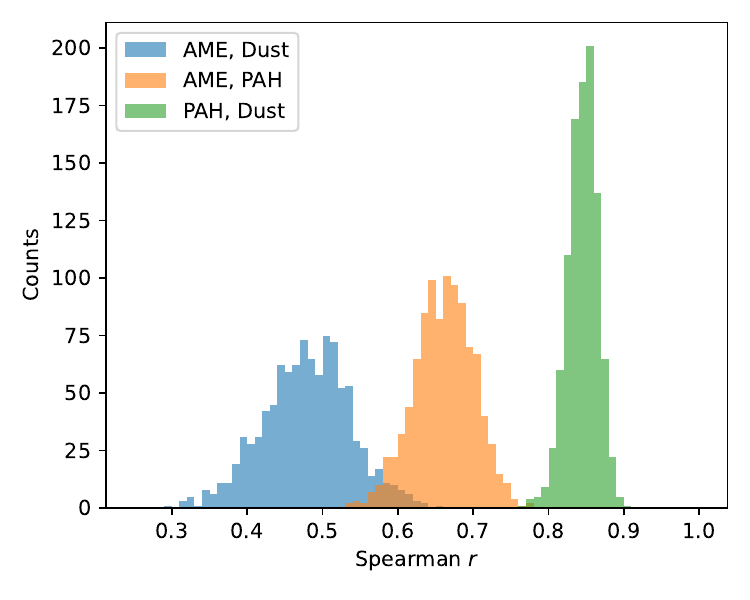"}
    \caption{Results of bootstrap resampling to estimate the uncertainty in measured correlation coefficients for the Perseus Molecular Cloud. The minimal overlap between the AME-PAH and AME-Dust correlation coefficient distributions shows that the preference for PAHs as a tracer of AME is statistically significant in this region.}
    \label{fig:bootstrap}
\end{figure}

We also obtain uncertainties on our correlation coefficients using a bootstrap resampling approach.
In this technique, the pixels in a given region are resampled with replacement several times and the correlation coefficients are recorded.
The distribution of correlation coefficients provides an estimate on the uncertainty.
We use the \textsc{pymccorrelation}\footnote{\url{https://github.com/privong/pymccorrelation/tree/pymccorr}} Python package to do this \citep{Curran2014, Privon2020}.
An example of the estimated distributions in the correlation coefficients is shown for the Perseus Molecular Cloud in Figure \ref{fig:bootstrap}.
The clear separation between the AME-PAH and AME-Dust correlation coefficient distributions shows that the difference is statistically significant in this region.

After obtaining uncertainties on the correlation coefficients, we can determine which regions prefer either thermal dust emission or PAHs with significance $\sigma \geq 2$.
For each source, the significance at which PAHs are preferred over dust may be calculated as 
\begin{equation}
	\eta_{\mathrm{pref}} = \frac{r_{\mathrm{AP}} - r_{\mathrm{AD}}}{\sqrt{\sigma_{\mathrm{AP}}^2 + \sigma_{\mathrm{AD}}^2}}
	\label{eqn:sig_pref}
\end{equation}
where $r_{\mathrm{AP}}$ and $r_{\mathrm{AD}}$ are the correlation coefficients between AME vs. PAHs and between AME vs. thermal dust emission, respectively, and $\sigma_{\mathrm{AP}}$ and $\sigma_{\mathrm{AD}}$ are the corresponding uncertainties inferred from bootstrapping.
The absolute value of $\eta_{\mathrm{pref}}$ gives the strength of the preference and a negative value indicates that dust is the preferred tracer.

Table \ref{tab:sources} lists the results from the Spearman correlation analysis for all 98 AME sources.
Columns 3--5 list the Spearman rank correlation cofficients for the correlation between AME and PAHs ($r_{\mathrm{AP}}$), AME and thermal dust emission ($r_{\mathrm{AD}}$), and PAHs and thermal dust emission ($r_{\mathrm{PD}}$).
Columns 6--8 show the associated correlation coefficient uncertainties from the bootstrap analysis.
The final column shows the significance at which PAHs are preferred over thermal dust emission as a tracer of AME.
The data are sorted in descending order of the significance of the AME detection in the \textit{Planck} analysis.

\startlongtable
\begin{deluxetable*}{cCCCCCCC}
\tablehead{
    \colhead{Source} & \colhead{$r_{\mathrm{AP}}$} & \colhead{$r_{\mathrm{AD}}$} & \colhead{$r_{\mathrm{PD}}$} & \colhead{$\sigma_{\mathrm{AP}}$} & \colhead{$\sigma_{\mathrm{AD}}$} & \colhead{$\sigma_{\mathrm{PD}}$} & \colhead{$\eta_{\mathrm{pref}}$}
}
\tablecaption{Spearman correlation coefficients and uncertainties for the \textit{Planck} AME sources.}
\startdata
	\textbf{G353.05+16.90} & 0.885 & 0.623 & 0.577 & 0.015 & 0.041 & 0.050 & 6.03 \\ 
	\textbf{G219.18$\mathbf{-}$08.93} & 0.421 & 0.345 & 0.286 & 0.047 & 0.052 & 0.056 & 1.08 \\ 
	\textbf{G005.40+36.50} & 0.022 & 0.638 & 0.155 & 0.063 & 0.039 & 0.062 & -8.32 \\ 
	\textbf{G160.26$\mathbf{-}$18.62} & 0.662 & 0.481 & 0.848 & 0.041 & 0.055 & 0.019 & 2.66 \\ 
	G158.40$-$20.60 & 0.476 & -0.194 & 0.253 & 0.045 & 0.057 & 0.060 & 9.25 \\ 
	\textbf{G004.24+18.09} & 0.339 & 0.563 & 0.374 & 0.054 & 0.044 & 0.058 & -3.21 \\ 
	G213.71$-$12.60 & 0.361 & 0.259 & 0.563 & 0.055 & 0.060 & 0.047 & 1.26 \\ 
	\textbf{G234.20$\mathbf{-}$00.20} & 0.428 & 0.825 & 0.436 & 0.056 & 0.020 & 0.052 & -6.69 \\ 
	\textbf{G247.60$\mathbf{-}$12.40} & 0.400 & 0.469 & 0.422 & 0.062 & 0.054 & 0.053 & -0.84 \\ 
	\textbf{G355.63+20.52} & 0.822 & 0.583 & 0.649 & 0.018 & 0.045 & 0.039 & 4.91 \\ 
	\textbf{G180.80+04.30} & 0.050 & 0.519 & 0.094 & 0.060 & 0.050 & 0.065 & -6.02 \\ 
	\textbf{G192.34$\mathbf{-}$11.37} & 0.756 & 0.857 & 0.700 & 0.036 & 0.016 & 0.042 & -2.54 \\ 
	\textbf{G231.83$\mathbf{-}$02.00} & 0.399 & 0.832 & 0.490 & 0.050 & 0.019 & 0.048 & -8.06 \\ 
	\textbf{G353.97+15.79} & 0.830 & 0.750 & 0.734 & 0.022 & 0.022 & 0.031 & 2.56 \\ 
	\textbf{G259.30$\mathbf{-}$13.50} & 0.647 & 0.608 & 0.630 & 0.032 & 0.039 & 0.036 & 0.79 \\ 
	G107.20+05.20 & 0.782 & 0.772 & 0.797 & 0.030 & 0.028 & 0.026 & 0.22 \\ 
	\textbf{G239.40$\mathbf{-}$04.70} & 0.666 & 0.911 & 0.681 & 0.039 & 0.010 & 0.034 & -6.03 \\ 
	G253.80$-$00.20 & 0.516 & 0.833 & 0.668 & 0.043 & 0.023 & 0.033 & -6.44 \\ 
	\textbf{G142.35+01.35} & 0.573 & 0.617 & 0.803 & 0.048 & 0.045 & 0.024 & -0.66 \\ 
	G218.05$-$00.38 & 0.343 & 0.603 & 0.650 & 0.054 & 0.053 & 0.037 & -3.44 \\ 
	\textbf{G293.35$\mathbf{-}$24.47} & 0.770 & 0.876 & 0.746 & 0.026 & 0.013 & 0.033 & -3.58 \\ 
	\textbf{G133.27+09.05} & 0.375 & 0.219 & 0.473 & 0.046 & 0.054 & 0.049 & 2.19 \\ 
	G030.77$-$00.03 & 0.933 & 0.975 & 0.935 & 0.009 & 0.004 & 0.009 & -4.38 \\ 
	\textbf{G023.47+08.19} & 0.245 & 0.445 & 0.550 & 0.056 & 0.047 & 0.044 & -2.75 \\ 
	\textbf{G203.24+02.08} & 0.281 & 0.600 & 0.713 & 0.059 & 0.043 & 0.038 & -4.37 \\ 
	G062.98+00.05 & 0.805 & 0.903 & 0.768 & 0.020 & 0.010 & 0.030 & -4.38 \\ 
	G190.00+00.46 & 0.577 & 0.731 & 0.575 & 0.043 & 0.033 & 0.046 & -2.83 \\ 
	\textbf{G201.62+01.63} & 0.353 & 0.224 & 0.701 & 0.051 & 0.056 & 0.035 & 1.70 \\ 
	\textbf{G204.70$\mathbf{-}$11.80} & 0.342 & 0.472 & 0.464 & 0.056 & 0.054 & 0.050 & -1.67 \\ 
	G059.42$-$00.21 & 0.854 & 0.949 & 0.821 & 0.022 & 0.007 & 0.026 & -4.19 \\ 
	G320.27$-$00.27 & 0.618 & 0.912 & 0.746 & 0.037 & 0.013 & 0.028 & -7.53 \\ 
	G305.27+00.15 & 0.775 & 0.902 & 0.810 & 0.033 & 0.014 & 0.028 & -3.57 \\ 
	G098.00+01.47 & 0.346 & 0.434 & 0.500 & 0.052 & 0.053 & 0.051 & -1.18 \\ 
	\textbf{G344.75+23.97} & 0.636 & 0.518 & 0.630 & 0.040 & 0.043 & 0.039 & 2.02 \\ 
	G045.47+00.06 & 0.895 & 0.965 & 0.883 & 0.014 & 0.004 & 0.013 & -4.70 \\ 
	G173.62+02.79 & 0.500 & 0.796 & 0.600 & 0.044 & 0.020 & 0.041 & -6.19 \\ 
	\textbf{G211.98$\mathbf{-}$01.17} & 0.437 & 0.615 & 0.521 & 0.052 & 0.042 & 0.048 & -2.65 \\ 
	G043.20$-$00.10 & 0.872 & 0.978 & 0.858 & 0.013 & 0.003 & 0.014 & -8.12 \\ 
	\textbf{G017.00+00.85} & 0.676 & 0.910 & 0.595 & 0.033 & 0.011 & 0.042 & -6.82 \\ 
	\textbf{G351.31+17.28} & 0.904 & 0.848 & 0.792 & 0.015 & 0.023 & 0.032 & 2.07 \\ 
	\textbf{G318.49$\mathbf{-}$04.28} & 0.645 & 0.812 & 0.801 & 0.036 & 0.025 & 0.021 & -3.81 \\ 
	\textbf{G160.60$\mathbf{-}$12.05} & 0.150 & 0.539 & 0.684 & 0.060 & 0.046 & 0.031 & -5.19 \\ 
	G317.51$-$00.11 & 0.812 & 0.968 & 0.844 & 0.016 & 0.004 & 0.018 & -9.46 \\ 
	G345.40$-$00.94 & 0.726 & 0.951 & 0.759 & 0.028 & 0.008 & 0.025 & -7.61 \\ 
	G182.36+00.22 & 0.517 & 0.830 & 0.626 & 0.047 & 0.022 & 0.042 & -5.95 \\ 
	G282.02$-$01.16 & 0.892 & 0.941 & 0.853 & 0.014 & 0.008 & 0.019 & -3.05 \\ 
	G192.60$-$00.06 & 0.427 & 0.589 & 0.545 & 0.052 & 0.047 & 0.044 & -2.30 \\ 
	G166.44$-$24.08 & 0.214 & 0.448 & 0.494 & 0.060 & 0.054 & 0.047 & -2.90 \\ 
	G311.94+00.12 & 0.944 & 0.975 & 0.923 & 0.008 & 0.005 & 0.011 & -3.26 \\ 
	G243.16+00.42 & 0.181 & 0.765 & 0.304 & 0.047 & 0.028 & 0.055 & -10.71 \\ 
	G053.63+00.19 & 0.735 & 0.928 & 0.797 & 0.031 & 0.008 & 0.025 & -6.06 \\ 
	G110.25+02.58 & 0.629 & 0.851 & 0.684 & 0.040 & 0.018 & 0.035 & -5.06 \\ 
	G010.19$-$00.32 & 0.477 & 0.931 & 0.377 & 0.058 & 0.010 & 0.063 & -7.74 \\ 
	G037.79$-$00.11 & 0.776 & 0.966 & 0.741 & 0.028 & 0.005 & 0.029 & -6.77 \\ 
	G343.48$-$00.04 & 0.692 & 0.928 & 0.769 & 0.037 & 0.019 & 0.027 & -5.65 \\ 
	G260.50+00.40 & 0.568 & 0.813 & 0.616 & 0.041 & 0.022 & 0.042 & -5.23 \\ 
	G298.60$-$00.20 & 0.810 & 0.943 & 0.818 & 0.028 & 0.008 & 0.027 & -4.57 \\ 
	G353.16+00.74 & 0.595 & 0.881 & 0.649 & 0.040 & 0.021 & 0.043 & -6.39 \\ 
	G355.44+00.11 & 0.200 & 0.911 & 0.270 & 0.076 & 0.021 & 0.074 & -9.03 \\ 
	G336.90+00.00 & 0.844 & 0.979 & 0.866 & 0.019 & 0.003 & 0.018 & -7.10 \\ 
	G012.80$-$00.19 & 0.672 & 0.949 & 0.641 & 0.039 & 0.010 & 0.044 & -6.84 \\ 
	G075.81+00.39 & 0.847 & 0.829 & 0.843 & 0.019 & 0.023 & 0.018 & 0.61 \\ 
	G322.15+00.61 & 0.638 & 0.951 & 0.730 & 0.044 & 0.009 & 0.033 & -7.02 \\ 
	G102.88$-$00.69 & 0.797 & 0.931 & 0.757 & 0.026 & 0.008 & 0.028 & -4.99 \\ 
	G327.30$-$00.50 & 0.845 & 0.946 & 0.919 & 0.019 & 0.007 & 0.012 & -4.91 \\ 
	G284.30$-$00.30 & 0.860 & 0.869 & 0.908 & 0.015 & 0.016 & 0.013 & -0.42 \\ 
	G270.27+00.84 & 0.756 & 0.894 & 0.787 & 0.033 & 0.014 & 0.031 & -3.84 \\ 
	G351.65$-$01.23 & 0.846 & 0.905 & 0.901 & 0.019 & 0.020 & 0.014 & -2.13 \\ 
	G287.48$-$00.63 & 0.675 & 0.747 & 0.762 & 0.039 & 0.034 & 0.029 & -1.39 \\ 
	G208.80$-$02.65 & 0.656 & 0.733 & 0.763 & 0.041 & 0.031 & 0.027 & -1.52 \\ 
	G294.98$-$01.71 & 0.676 & 0.957 & 0.671 & 0.038 & 0.007 & 0.036 & -7.22 \\ 
	G035.20$-$01.74 & 0.941 & 0.977 & 0.954 & 0.012 & 0.003 & 0.009 & -2.93 \\ 
	G267.95$-$01.06 & 0.391 & 0.504 & 0.830 & 0.053 & 0.046 & 0.024 & -1.63 \\ 
	G291.63$-$00.52 & 0.757 & 0.863 & 0.809 & 0.030 & 0.025 & 0.023 & -2.69 \\ 
	G008.51$-$00.31 & 0.350 & 0.938 & 0.320 & 0.061 & 0.008 & 0.060 & -9.57 \\ 
	G015.06$-$00.69 & 0.811 & 0.922 & 0.827 & 0.027 & 0.015 & 0.025 & -3.58 \\ 
	G061.47+00.11 & 0.761 & 0.902 & 0.810 & 0.033 & 0.012 & 0.027 & -4.01 \\ 
	G209.01$-$19.38 & 0.822 & 0.688 & 0.821 & 0.030 & 0.040 & 0.025 & 2.65 \\ 
	G123.13$-$06.27 & 0.369 & 0.544 & 0.470 & 0.058 & 0.046 & 0.051 & -2.37 \\ 
	G071.59+02.85 & 0.925 & 0.967 & 0.930 & 0.007 & 0.006 & 0.008 & -4.32 \\ 
	G274.01$-$01.15 & 0.501 & 0.766 & 0.605 & 0.047 & 0.032 & 0.042 & -4.64 \\ 
	G093.02+02.76 & 0.544 & 0.705 & 0.653 & 0.046 & 0.033 & 0.040 & -2.84 \\ 
	G109.01+00.00 & 0.603 & 0.563 & 0.695 & 0.040 & 0.052 & 0.036 & 0.62 \\ 
	G133.74+01.22 & 0.688 & 0.717 & 0.773 & 0.033 & 0.030 & 0.028 & -0.66 \\ 
	G151.62$-$00.28 & 0.238 & 0.746 & 0.318 & 0.061 & 0.030 & 0.059 & -7.45 \\ 
	G081.59+00.01 & 0.733 & 0.894 & 0.693 & 0.050 & 0.013 & 0.054 & -3.14 \\ 
	G351.29+00.68 & 0.649 & 0.893 & 0.766 & 0.034 & 0.020 & 0.028 & -6.18 \\ 
	G265.15+01.45 & 0.506 & 0.658 & 0.781 & 0.049 & 0.047 & 0.025 & -2.26 \\ 
	G173.56$-$01.76 & 0.437 & 0.524 & 0.560 & 0.049 & 0.044 & 0.040 & -1.33 \\ 
	G348.73$-$00.75 & 0.683 & 0.952 & 0.667 & 0.043 & 0.011 & 0.046 & -6.13 \\ 
	G094.47$-$01.53 & 0.770 & 0.961 & 0.772 & 0.032 & 0.005 & 0.032 & -5.97 \\ 
	G099.60+03.70 & 0.392 & 0.696 & 0.506 & 0.056 & 0.037 & 0.052 & -4.52 \\ 
	G040.52+02.53 & 0.976 & 0.970 & 0.953 & 0.003 & 0.006 & 0.007 & 0.87 \\ 
	G028.79+03.49 & 0.615 & 0.628 & 0.706 & 0.035 & 0.038 & 0.034 & -0.26 \\ 
	G068.16+01.02 & 0.590 & 0.852 & 0.432 & 0.038 & 0.018 & 0.054 & -6.29 \\ 
	G076.38$-$00.62 & 0.824 & 0.918 & 0.850 & 0.022 & 0.011 & 0.017 & -3.89 \\ 
	G118.09+04.96 & 0.593 & 0.614 & 0.668 & 0.044 & 0.047 & 0.036 & -0.33 \\ 
	G289.80$-$01.15 & 0.669 & 0.794 & 0.731 & 0.042 & 0.038 & 0.032 & -2.20 \\ 
\enddata
\tablecomments{
The table is sorted according to the \textit{Planck} detection significance.
Boldface is used to indicate sources which were flagged in the \textit{Planck} analysis as highly significant; note that some sources with an SNR greater than 5 are excluded due to potentially significant contributions from ultra-compact HII regions.
Columns $r_{\mathrm{AP}}$, $r_{\mathrm{AD}}$, and $r_{\mathrm{PD}}$ give the Spearman rank correlation coefficients for the correlations between AME and PAH, AME and dust, and PAH and dust, respectively.
Columns $\sigma_{\mathrm{AP}}$, $\sigma_{\mathrm{AD}}$, and $\sigma_{\mathrm{PD}}$ give the estimated uncertainties in these correlation coefficients as obtained through bootstrap resampling (see text).
The final column gives the significance at which one tracer of AME is preferred over the other.
The magnitude indicates the strength of the preference, while a positive value indicates a preference for PAHs and a negative value indicates a preference for dust.
Of the 98 sources considered, nine were better spatially correlated with small PAHs than with thermal dust emission from larger grains with a statistical significance of $\sigma > 2$.}
\label{tab:sources}
\end{deluxetable*}

\begin{figure}
    \centering
    \includegraphics[width=\columnwidth]{"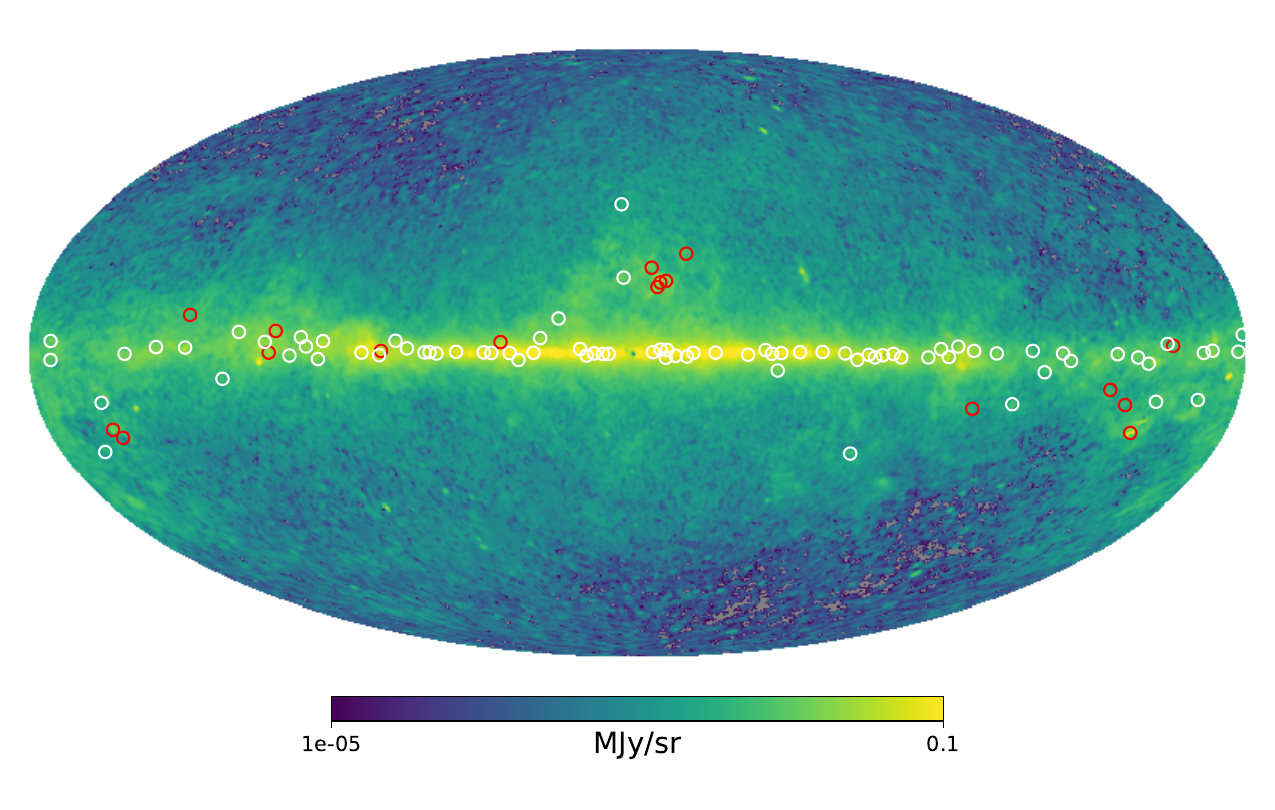"}
    \caption{
    AME sources identified by \textit{Planck}.
	The background map shows the total AME at 30 GHz
	calculated from a \textit{Planck} fit to a two-component AME model \citep{Planck-Collaboration2016d}.
    The colorbar uses a logarithmic scale.
    All 98 considered sources are indicated by circles, with red circles indicating those for which $r_{\mathrm{AP}} > r_{\mathrm{AD}}$ and white circles indicating sources for which $r_{\mathrm{AP}} < r_{\mathrm{AD}}$.
    }
    \label{fig:AMEsources}
\end{figure}

\begin{figure}
    \centering
    \includegraphics[width=\columnwidth]{"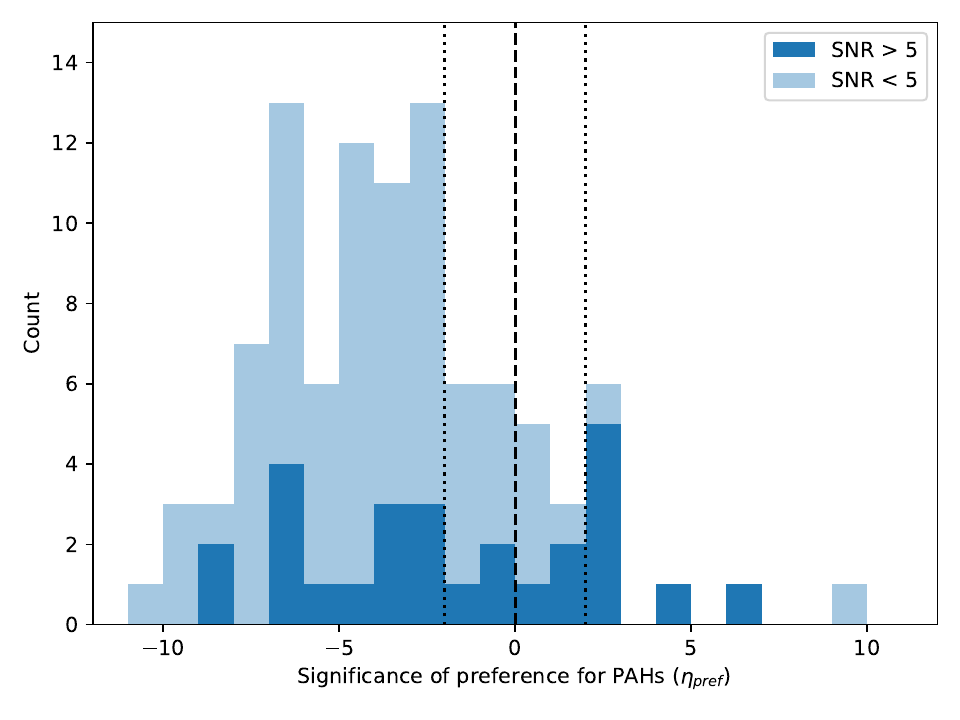"}
    \caption{A histogram of the significance at which PAHs are the preferred tracer for AME. The absolute magnitude indicates the statistical significance, and a positive (negative) value indicates that PAHs (large dust grains) are the best tracer. A notable fraction of sources prefer PAHs over thermal dust emission, and this becomes more pronounced when considering only those flagged by the \textit{Planck} team as the most significant AME detections (darker shade).}
    \label{fig:sig_pref_PAH_hist}
\end{figure}

Figure \ref{fig:AMEsources} shows the map of AME at 30 GHz obtained using the \textit{Planck} model parameters along with the locations of the 98 sources.
Sources which are better correlated with PAHs than with thermal dust emission are shown in red.
The PAH-correlated sources indicated in red include the Perseus Molecular Cloud as well as $\rho$-Ophiuchus, which has a strong preference for PAHs.
Notably, many of the PAH-correlated sources lie off of the Galactic plane, where free-free emission is less likely to complicate the separation of AME.
Of the 98 sources considered here, 17\% prefer PAHs as the AME tracer over thermal dust emission.
Nine prefer PAHs with a significance of $\eta_{\mathrm{pref}} \geq 2$ while 69 significantly prefer thermal dust emission, leading to 12\% of sources with $|\eta_{\mathrm{pref}}| \geq 2$ having a strong preference for PAHs.
If we restrict our analysis to consider only the 27 regions identified in the \textit{Planck} analysis as having strong AME detections, we find that ten are better correlated with PAH emission (37\%) while 17 prefer dust.
Further restricting these data to eliminate sources with preferences smaller than 2$\sigma$, we find 7 remaining AME sources better correlated with PAH emission (33\%) while 14 are better correlated with thermal dust emission (67\%).
We summarize results in Table \ref{tab:results} and Figure \ref{fig:sig_pref_PAH_hist}.

\begin{deluxetable*}{cccc}
\tablehead{
    \colhead{Subset of Sources}   &   \colhead{Prefer PAHs}   &   \colhead{Prefer Dust}   &   \colhead{\% PAHs}
}
\tablecaption{Preferred tracers of AME}
\startdata
    All Sources 										&   17  &   81  &   17\% \\
    \textit{Planck} Flagged Sources 						&   10  &   17  &   37\% \\
    Sources with a strong preference 						&   9   &   69  &   12\% \\
    \textit{Planck} Flagged Sources with a strong preference 	&   7   &   14  &   33\% \\
\enddata
\tablecomments{
Preferred tracers of AME for different subsets of the source sample. Looking at all sources the majority seem to favor thermal dust emission, however a notable subset prefer PAHs. The distribution is more balanced between the two when only considering the sources identified as highly significant AME detections in the \textit{Planck} analysis. If we only look at sources that have a significant preference for one tracer over the other, the number of sources is reduced but the relative balance between how many prefer dust vs. PAHs is not significantly affected. These results are also shown in Figure \ref{fig:sig_pref_PAH_hist}.
}
\label{tab:results}
\end{deluxetable*}

\section{Discussion and Summary}
\label{sec:conclusions}

Here we have analyzed a large number of AME sources to discern the possible connection between AME, dust, and PAHs.
Over this sample, we find that neither the 3.3 $\mu$m PAH feature nor the thermal dust emission at 857 GHz is sufficient to serve as a tracer of AME on its own.
While most sources prefer thermal dust emission, a notable subset of 17\% of the considered regions prefer emission from small PAHs.
This increases to 37\% of sources if we restrict the sample to only those with high detection significance.
This suggests that there is evidence in favor of both thermal emission from larger dust grains and PAH emission as tracers of AME.
This may point to more nuanced influences from the surrounding environment, such as unresolved small-scale structure, which may be needed to understand the underlying mechanism behind AME.
It is also possible that multiple components of the ISM are contributing to AME.

\cite{Poidevin2023} analyze 42 of the AME sources identified by \cite{Planck-Collaboration2014a} with the addition of low-frequency data from the QUIJOTE-MFI survey \citep{Rubino-Martin2023}, allowing a more reliable separation between AME and free-free emission.
If we replace the AME detection significances from \textit{Planck} with those from QUIJOTE for all applicable sources, the number of sources with significant AME detections increases from 27 to 43 and the number of significant sources which favor PAHs increases proportionally from 10 to 17 (39\% favoring PAHs).

We note that beam correlations can have an impact on this analysis, as the \textit{Planck} AME map has a resolution of one degree while the analysis is performed at $N_{\mathrm{side}} = 256$, which corresponds to a pixel size of approximately 0.23 degrees.
For Nyquist sampling the pixel size should be half the beam size.
To assess the possible effect of correlated pixels, we convolved the maps with the \textit{Planck} beam and degraded them to $N_{\mathrm{side}} = 128$ and repeated the correlation analysis.
We found that this did not significantly impact our results on correlation coefficients or on the number of sources which show a stronger correlation between AME and PAHs than between AME and thermal dust emission.

Finally, We emphasize that the AME map is likely biased due to complications in separating it from other components, such as free-free emission, as well as limitations in the frequency range of available data used to fit the AME spectrum.
The increased fraction of PAH-correlated sources at high latitudes, where AME detections are generally more robust, may indicate that PAH emission will prove to be a better tracer as higher angular resolution data becomes available across a broader frequency range.
PAH emission mechanisms could also be affected by varying interstellar conditions, artificially suppressing the correlation between PAH emission and AME.
Higher resolution maps of PAH emission, such as observations of the 3.3 $\mu$m emission feature that will be obtained by SPHEREx \citep{Dore2014}, may also be crucial to determining the viability of PAHs as a carrier of AME.

This work supplements previous studies by considering a larger sample of AME regions, allowing us to obtain a statistical understanding of how often AME prefers dust over PAHs as a tracer and vice versa.
It has resulted in the identification of several regions in which AME is better correlated with PAHs than with thermal dust emission.
To further improve our understanding of AME tracers, future work will entail a similar analysis of the DIRBE data over the full sky, allowing us to probe diffuse regions at high Galactic latitude in addition to these compact sources.

\begin{acknowledgments}
This work was funded by NASA ADAP award number 80NSSC24K0623.
This research was carried out in part at the Jet Propulsion Laboratory, California Institute of Technology, under a contract with the National Aeronautics and Space Administration.
We acknowledge the use of data provided by the Centre d'Analyse de Donn\'ees Etendues (CADE), a service of IRAP-UPS/CNRS (\url{http://cade.irap.omp.eu}, \cite{Paradis2012}).
We would also like to thank Dylan Par\'e for the many helpful suggestions.
\end{acknowledgments}

\bibliography{bib}{}

\begin{thebibliography}{}
\expandafter\ifx\csname natexlab\endcsname\relax\def\natexlab#1{#1}\fi
\providecommand{\url}[1]{\href{#1}{#1}}
\providecommand{\dodoi}[1]{doi:~\href{http://doi.org/#1}{\nolinkurl{#1}}}
\providecommand{\doeprint}[1]{\href{http://ascl.net/#1}{\nolinkurl{http://ascl.net/#1}}}
\providecommand{\doarXiv}[1]{\href{https://arxiv.org/abs/#1}{\nolinkurl{https://arxiv.org/abs/#1}}}

\bibitem[{Abitbol {et~al.}(2017)Abitbol, Chluba, Hill, \&
  Johnson}]{Abitbol2017a}
Abitbol, M.~H., Chluba, J., Hill, J.~C., \& Johnson, B.~R. 2017, Monthly
  Notices of the Royal Astronomical Society, 471, 1126

\bibitem[{Ali-Ha{\"\i}moud {et~al.}(2009)Ali-Ha{\"\i}moud, Hirata, \&
  Dickinson}]{Ali-Haimoud2009}
Ali-Ha{\"\i}moud, Y., Hirata, C.~M., \& Dickinson, C. 2009, Monthly Notices of
  the Royal Astronomical Society, 395, 1055

\bibitem[{{Arce-Tord} {et~al.}(2020){Arce-Tord}, {Vidal}, {Casassus},
  {C{\'a}rcamo}, {Dickinson}, {Hensley}, {G{\'e}nova-Santos}, {Bond}, {Jones},
  {Readhead}, {Taylor}, \& {Zensus}}]{Arce-Tord2020}
{Arce-Tord}, C., {Vidal}, M., {Casassus}, S., {et~al.} 2020, \mnras, 495, 3482

\bibitem[{Bell {et~al.}(2019)Bell, Onaka, Galliano, Wu, Doi, Kaneda, Ishihara,
  \& Giard}]{Bell2019}
Bell, A.~C., Onaka, T., Galliano, F., {et~al.} 2019, Publications of the
  Astronomical Society of Japan, 71

\bibitem[{{Casassus} {et~al.}(2021){Casassus}, {Vidal}, {Arce-Tord},
  {Dickinson}, {White}, {Burton}, {Indermuehle}, \& {Hensley}}]{Casassus2021}
{Casassus}, S., {Vidal}, M., {Arce-Tord}, C., {et~al.} 2021, \mnras, 502, 589

\bibitem[{Chuss {et~al.}(2022)Chuss, Hensley, Kogut, Guerra, Nofi, \&
  Siah}]{Chuss2022}
Chuss, D.~T., Hensley, B.~S., Kogut, A.~J., {et~al.} 2022, The Astrophysical
  Journal, 940, 59

\bibitem[{{Curran}(2014)}]{Curran2014}
{Curran}, P.~A. 2014, arXiv e-prints, arXiv:1411.3816

\bibitem[{{Dickinson} {et~al.}(2011){Dickinson}, {Peel}, \&
  {Vidal}}]{Dickinson2011}
{Dickinson}, C., {Peel}, M., \& {Vidal}, M. 2011, \mnras, 418, L35

\bibitem[{Dickinson {et~al.}(2018)Dickinson, Ali-Ha\"{\i}moud, Barr,
  Battistelli, Bell, Bernstein, Casassus, Cleary, Draine, G\'{e}nova-Santos,
  Harper, Hensley, Hill-Valler, Hoang, Israel, Jew, Lazarian, Leahy, Leech,
  L\'{o}pez-Caraballo, McDonald, Murphy, Onaka, Paladini, Peel, Perrott,
  Poidevin, Readhead, {n}o Mart\'{i}n, Taylor, Tibbs, Todorovi\'{c}, \&
  Vidal}]{Dickinson2018}
Dickinson, C., Ali-Ha\"{\i}moud, Y., Barr, A., {et~al.} 2018, New Astronomy
  Reviews, 80, 1

\bibitem[{{Dor{\'e}} {et~al.}(2014){Dor{\'e}}, {Bock}, {Ashby}, {Capak},
  {Cooray}, {de Putter}, {Eifler}, {Flagey}, {Gong}, {Habib}, {Heitmann},
  {Hirata}, {Jeong}, {Katti}, {Korngut}, {Krause}, {Lee}, {Masters},
  {Mauskopf}, {Melnick}, {Mennesson}, {Nguyen}, {{\"O}berg}, {Pullen},
  {Raccanelli}, {Smith}, {Song}, {Tolls}, {Unwin}, {Venumadhav}, {Viero},
  {Werner}, \& {Zemcov}}]{Dore2014}
{Dor{\'e}}, O., {Bock}, J., {Ashby}, M., {et~al.} 2014, arXiv e-prints,
  arXiv:1412.4872

\bibitem[{Draine \& Hensley(2013)}]{Draine2013}
Draine, B.~T., \& Hensley, B. 2013, The Astrophysical Journal, 765, 159

\bibitem[{Draine \& Lazarian(1998)}]{Draine1998a}
Draine, B.~T., \& Lazarian, A. 1998, The Astrophysical Journal, 508, 157

\bibitem[{Draine \& Lazarian(1999)}]{Draine1999}
---. 1999, The Astrophysical Journal, 512, 740

\bibitem[{{Eriksen} {et~al.}(2008){Eriksen}, {Jewell}, {Dickinson}, {Banday},
  {G{\'o}rski}, \& {Lawrence}}]{Eriksen2008}
{Eriksen}, H.~K., {Jewell}, J.~B., {Dickinson}, C., {et~al.} 2008, \apj, 676,
  10

\bibitem[{{G{\'e}nova-Santos} {et~al.}(2015){G{\'e}nova-Santos},
  {Rubi{\~n}o-Mart{\'\i}n}, {Rebolo}, {Pel{\'a}ez-Santos},
  {L{\'o}pez-Caraballo}, {Harper}, {Watson}, {Ashdown}, {Barreiro},
  {Casaponsa}, {Dickinson}, {Diego}, {Fern{\'a}ndez-Cobos}, {Grainge},
  {Guti{\'e}rrez}, {Herranz}, {Hoyland}, {Lasenby}, {L{\'o}pez-Caniego},
  {Mart{\'\i}nez-Gonz{\'a}lez}, {McCulloch}, {Melhuish}, {Piccirillo},
  {Perrott}, {Poidevin}, {Razavi-Ghods}, {Scott}, {Titterington}, {Tramonte},
  {Vielva}, \& {Vignaga}}]{Genova-Santos2015}
{G{\'e}nova-Santos}, R., {Rubi{\~n}o-Mart{\'\i}n}, J.~A., {Rebolo}, R.,
  {et~al.} 2015, \mnras, 452, 4169

\bibitem[{{G{\'e}nova-Santos} {et~al.}(2017){G{\'e}nova-Santos},
  {Rubi{\~n}o-Mart{\'\i}n}, {Pel{\'a}ez-Santos}, {Poidevin}, {Rebolo},
  {Vignaga}, {Artal}, {Harper}, {Hoyland}, {Lasenby},
  {Mart{\'\i}nez-Gonz{\'a}lez}, {Piccirillo}, {Tramonte}, \&
  {Watson}}]{Genova-Santos2017}
{G{\'e}nova-Santos}, R., {Rubi{\~n}o-Mart{\'\i}n}, J.~A., {Pel{\'a}ez-Santos},
  A., {et~al.} 2017, \mnras, 464, 4107

\bibitem[{Gorski {et~al.}(2005)Gorski, Hivon, Banday, Wandelt, Hansen,
  Reinecke, \& Bartelmann}]{Gorski2005}
Gorski, K.~M., Hivon, E., Banday, A.~J., {et~al.} 2005, The Astrophysical
  Journal, 622, 759

\bibitem[{{Harper} {et~al.}(2025){Harper}, {Dickinson}, {Cleary}, {Hensley},
  {Hoerning}, {Paladini}, {Rennie}, {Cepeda-Arroita}, {Dunne}, {Eriksen},
  {Gundersen}, {Ihle}, {Lunde}, {Ricci}, {Stil}, {Stutzer}, {Taylor}, \&
  {Wehus}}]{Harper2025}
{Harper}, S.~E., {Dickinson}, C., {Cleary}, K.~A., {et~al.} 2025, \mnras, 536,
  2914

\bibitem[{Hensley \& Draine(2023)}]{Hensley2023}
Hensley, B.~S., \& Draine, B.~T. 2023, The Astrophysical Journal, 948, 55

\bibitem[{Hensley {et~al.}(2016)Hensley, Draine, \& Meisner}]{Hensley2016}
Hensley, B.~S., Draine, B.~T., \& Meisner, A.~M. 2016, The Astrophysical
  Journal, 827, 45

\bibitem[{{Herman} {et~al.}(2023){Herman}, {Hensley}, {Andersen}, {Aurlien},
  {Banerji}, {Bersanelli}, {Bertocco}, {Brilenkov}, {Carbone}, {Colombo},
  {Eriksen}, {Foss}, {Fuskeland}, {Galeotta}, {Galloway}, {Gerakakis},
  {Gjerl{\o}w}, {Iacobellis}, {Ieronymaki}, {Ihle}, {Jewell}, {Karakci},
  {Keih{\"a}nen}, {Keskitalo}, {Maggio}, {Maino}, {Maris}, {Paradiso},
  {Partridge}, {Reinecke}, {Suur-Uski}, {Svalheim}, {Tavagnacco}, {Thommesen},
  {Wehus}, \& {Zacchei}}]{Herman2023}
{Herman}, D., {Hensley}, B., {Andersen}, K.~J., {et~al.} 2023, \aap, 675, A15

\bibitem[{{Jones}(2009)}]{Jones2009}
{Jones}, A.~P. 2009, \aap, 506, 797

\bibitem[{Kelsall {et~al.}(1998)Kelsall, Weiland, Franz, Reach, Arendt, Dwek,
  Freudenreich, Hauser, Moseley, Odegard, Silverberg, \& Wright}]{Kelsall1998}
Kelsall, T., Weiland, J.~L., Franz, B.~A., {et~al.} 1998, The Astrophysical
  Journal, 508, 44

\bibitem[{{Kogut} {et~al.}(1996){Kogut}, {Banday}, {Bennett}, {Gorski},
  {Hinshaw}, \& {Reach}}]{Kogut1996a}
{Kogut}, A., {Banday}, A.~J., {Bennett}, C.~L., {et~al.} 1996, \apj, 460, 1

\bibitem[{Leitch {et~al.}(1997)Leitch, Readhead, Pearson, \&
  Myers}]{Leitch1997}
Leitch, E.~M., Readhead, A. C.~S., Pearson, T.~J., \& Myers, S.~T. 1997, The
  Astrophysical Journal, 486, L23

\bibitem[{{Li} \& {Draine}(2001)}]{Li2001}
{Li}, A., \& {Draine}, B.~T. 2001, \apj, 554, 778

\bibitem[{{Maddalena} \& {Morris}(1987)}]{Maddalena1987}
{Maddalena}, R.~J., \& {Morris}, M. 1987, \apj, 323, 179

\bibitem[{{Mathis} {et~al.}(1983){Mathis}, {Mezger}, \& {Panagia}}]{Mathis1983}
{Mathis}, J.~S., {Mezger}, P.~G., \& {Panagia}, N. 1983, \aap, 500, 259

\bibitem[{{Meny} {et~al.}(2007){Meny}, {Gromov}, {Boudet}, {Bernard},
  {Paradis}, \& {Nayral}}]{Meny2007}
{Meny}, C., {Gromov}, V., {Boudet}, N., {et~al.} 2007, \aap, 468, 171

\bibitem[{{Mezger} {et~al.}(1982){Mezger}, {Mathis}, \& {Panagia}}]{Mezger1982}
{Mezger}, P.~G., {Mathis}, J.~S., \& {Panagia}, N. 1982, \aap, 105, 372

\bibitem[{{Nashimoto} {et~al.}(2020){Nashimoto}, {Hattori}, {Poidevin}, \&
  {G{\'e}nova-Santos}}]{Nashimoto2020}
{Nashimoto}, M., {Hattori}, M., {Poidevin}, F., \& {G{\'e}nova-Santos}, R.
  2020, \apjl, 900, L40

\bibitem[{{Paradis} {et~al.}(2012){Paradis}, {Dobashi}, {Shimoikura},
  {Kawamura}, {Onishi}, {Fukui}, \& {Bernard}}]{Paradis2012}
{Paradis}, D., {Dobashi}, K., {Shimoikura}, T., {et~al.} 2012, \aap, 543, A103

\bibitem[{{Planck Collaboration} {et~al.}(2011){Planck Collaboration}, {Ade, P.
  A. R.}, {Aghanim, N.}, {Arnaud, M.}, {Ashdown, M.}, {Aumont, J.},
  {Baccigalupi, C.}, {Balbi, A.}, {Banday, A. J.}, {Barreiro, R. B.},
  {Bartlett, J. G.}, {Battaner, E.}, {Benabed, K.}, {Beno\^{\i}t, A.},
  {Bernard, J.-P.}, {Bersanelli, M.}, {Bhatia, R.}, {Bock, J. J.}, {Bonaldi,
  A.}, {Bond, J. R.}, {Borrill, J.}, {Bouchet, F. R.}, {Boulanger, F.},
  {Bucher, M.}, {Burigana, C.}, {Cabella, P.}, {Cappellini, B.}, {Cardoso,
  J.-F.}, {Casassus, S.}, {Catalano, A.}, {Cay\'on, L.}, {Challinor, A.},
  {Chamballu, A.}, {Chary, R.-R.}, {Chen, X.}, {Chiang, L.-Y.}, {Chiang, C.},
  {Christensen, P. R.}, {Clements, D. L.}, {Colombi, S.}, {Couchot, F.},
  {Coulais, A.}, {Crill, B. P.}, {Cuttaia, F.}, {Danese, L.}, {Davies, R. D.},
  {Davis, R. J.}, {de Bernardis, P.}, {de Gasperis, G.}, {de Rosa, A.}, {de
  Zotti, G.}, {Delabrouille, J.}, {Delouis, J.-M.}, {Dickinson, C.}, {Donzelli,
  S.}, {Dor\'e, O.}, {D\"orl, U.}, {Douspis, M.}, {Dupac, X.}, {Efstathiou,
  G.}, {En\ss{}lin, T. A.}, {Eriksen, H. K.}, {Finelli, F.}, {Forni, O.},
  {Frailis, M.}, {Franceschi, E.}, {Galeotta, S.}, {Ganga, K.},
  {G\'enova-Santos, R. T.}, {Giard, M.}, {Giardino, G.}, {Giraud-H\'eraud, Y.},
  {Gonz\'alez-Nuevo, J.}, {G\'orski, K. M.}, {Gratton, S.}, {Gregorio, A.},
  {Gruppuso, A.}, {Hansen, F. K.}, {Harrison, D.}, {Helou, G.},
  {Henrot-Versill\'e, S.}, {Herranz, D.}, {Hildebrandt, S. R.}, {Hivon, E.},
  {Hobson, M.}, {Holmes, W. A.}, {Hovest, W.}, {Hoyland, R. J.}, {Huffenberger,
  K. M.}, {Jaffe, T. R.}, {Jaffe, A. H.}, {Jones, W. C.}, {Juvela, M.},
  {Keih\"anen, E.}, {Keskitalo, R.}, {Kisner, T. S.}, {Kneissl, R.}, {Knox,
  L.}, {Kurki-Suonio, H.}, {Lagache, G.}, {L\"ahteenm\"aki, A.}, {Lamarre,
  J.-M.}, {Lasenby, A.}, {Laureijs, R. J.}, {Lawrence, C. R.}, {Leach, S.},
  {Leonardi, R.}, {Lilje, P. B.}, {Linden-V\o{}rnle, M.}, {L\'opez-Caniego,
  M.}, {Lubin, P. M.}, {Mac\'{\i}as-P\'erez, J. F.}, {MacTavish, C. J.},
  {Maffei, B.}, {Maino, D.}, {Mandolesi, N.}, {Mann, R.}, {Maris, M.},
  {Marshall, D. J.}, {Mart\'{\i}nez-Gonz\'alez, E.}, {Masi, S.}, {Matarrese,
  S.}, {Matthai, F.}, {Mazzotta, P.}, {McGehee, P.}, {Meinhold, P. R.},
  {Melchiorri, A.}, {Mendes, L.}, {Mennella, A.}, {Mitra, S.},
  {Miville-Desch\^enes, M.-A.}, {Moneti, A.}, {Montier, L.}, {Morgante, G.},
  {Mortlock, D.}, {Munshi, D.}, {Murphy, A.}, {Naselsky, P.}, {Natoli, P.},
  {Netterfield, C. B.}, {N\o{}rgaard-Nielsen, H. U.}, {Noviello, F.}, {Novikov,
  D.}, {Novikov, I.}, {O\'{}Dwyer, I. J.}, {Osborne, S.}, {Pajot, F.},
  {Paladini, R.}, {Partridge, B.}, {Pasian, F.}, {Patanchon, G.}, {Pearson, T.
  J.}, {Peel, M.}, {Perdereau, O.}, {Perotto, L.}, {Perrotta, F.}, {Piacentini,
  F.}, {Piat, M.}, {Plaszczynski, S.}, {Platania, P.}, {Pointecouteau, E.},
  {Polenta, G.}, {Ponthieu, N.}, {Poutanen, T.}, {Pr\'ezeau, G.}, {Procopio,
  P.}, {Prunet, S.}, {Puget, J.-L.}, {Reach, W. T.}, {Rebolo, R.}, {Reich, W.},
  {Reinecke, M.}, {Renault, C.}, {Ricciardi, S.}, {Riller, T.}, {Ristorcelli,
  I.}, {Rocha, G.}, {Rosset, C.}, {Rowan-Robinson, M.}, {Rubi\~no-Mart\'{\i}n,
  J. A.}, {Rusholme, B.}, {Sandri, M.}, {Santos, D.}, {Savini, G.}, {Scott,
  D.}, {Seiffert, M. D.}, {Shellard, P.}, {Smoot, G. F.}, {Starck, J.-L.},
  {Stivoli, F.}, {Stolyarov, V.}, {Stompor, R.}, {Sudiwala, R.}, {Sygnet,
  J.-F.}, {Tauber, J. A.}, {Terenzi, L.}, {Toffolatti, L.}, {Tomasi, M.},
  {Torre, J.-P.}, {Tristram, M.}, {Tuovinen, J.}, {Umana, G.}, {Valenziano,
  L.}, {Varis, J.}, {Verstraete, L.}, {Vielva, P.}, {Villa, F.}, {Vittorio,
  N.}, {Wade, L. A.}, {Wandelt, B. D.}, {Watson, R.}, {Wilkinson, A.}, {Ysard,
  N.}, {Yvon, D.}, {Zacchei, A.}, \& {Zonca, A.}}]{Planck-Collaboration2011}
{Planck Collaboration}, {Ade, P. A. R.}, {Aghanim, N.}, {et~al.} 2011, A\&A,
  536, A20

\bibitem[{{Planck Collaboration} {et~al.}(2014){Planck Collaboration}, {Ade, P.
  A. R.}, {Aghanim, N.}, {Alves, M. I. R.}, {Arnaud, M.}, {Atrio-Barandela,
  F.}, {Aumont, J.}, {Baccigalupi, C.}, {Banday, A. J.}, {Barreiro, R. B.},
  {Battaner, E.}, {Benabed, K.}, {Benoit-L\'evy, A.}, {Bernard, J.-P.},
  {Bersanelli, M.}, {Bielewicz, P.}, {Bobin, J.}, {Bonaldi, A.}, {Bond, J. R.},
  {Borrill, J.}, {Bouchet, F. R.}, {Boulanger, F.}, {Burigana, C.}, {Cardoso,
  J.-F.}, {Casassus, S.}, {Catalano, A.}, {Chamballu, A.}, {Chen, X.}, {Chiang,
  H. C.}, {Chiang, L.-Y.}, {Christensen, P. R.}, {Clements, D. L.}, {Colombi,
  S.}, {Colombo, L. P. L.}, {Couchot, F.}, {Crill, B. P.}, {Cuttaia, F.},
  {Danese, L.}, {Davies, R. D.}, {Davis, R. J.}, {de Bernardis, P.}, {de Rosa,
  A.}, {de Zotti, G.}, {Delabrouille, J.}, {D\'esert, F.-X.}, {Dickinson, C.},
  {Diego, J. M.}, {Donzelli, S.}, {Dor\'e, O.}, {Dupac, X.}, {En\ss{}lin, T.
  A.}, {Eriksen, H. K.}, {Finelli, F.}, {Forni, O.}, {Franceschi, E.},
  {Galeotta, S.}, {Ganga, K.}, {G\'enova-Santos, R. T.}, {Ghosh, T.}, {Giard,
  M.}, {Gonz\'alez-Nuevo, J.}, {G\'orski, K. M.}, {Gregorio, A.}, {Gruppuso,
  A.}, {Hansen, F. K.}, {Harrison, D. L.}, {Helou, G.},
  {Hern\'andez-Monteagudo, C.}, {Hildebrandt, S. R.}, {Hivon, E.}, {Hobson,
  M.}, {Hornstrup, A.}, {Jaffe, A. H.}, {Jaffe, T. R.}, {Jones, W. C.},
  {Keih\"anen, E.}, {Keskitalo, R.}, {Kneissl, R.}, {Knoche, J.}, {Kunz, M.},
  {Kurki-Suonio, H.}, {L\"ahteenm\"aki, A.}, {Lamarre, J.-M.}, {Lasenby, A.},
  {Lawrence, C. R.}, {Leonardi, R.}, {Liguori, M.}, {Lilje, P. B.},
  {Linden-V\o{}rnle, M.}, {L\'opez-Caniego, M.}, {Mac\'{\i}as-P\'erez, J. F.},
  {Maffei, B.}, {Maino, D.}, {Mandolesi, N.}, {Marshall, D. J.}, {Martin, P.
  G.}, {Mart\'{\i}nez-Gonz\'alez, E.}, {Masi, S.}, {Massardi, M.}, {Matarrese,
  S.}, {Mazzotta, P.}, {Meinhold, P. R.}, {Melchiorri, A.}, {Mendes, L.},
  {Mennella, A.}, {Migliaccio, M.}, {Miville-Desch\^enes, M.-A.}, {Moneti, A.},
  {Montier, L.}, {Morgante, G.}, {Mortlock, D.}, {Munshi, D.}, {Naselsky, P.},
  {Nati, F.}, {Natoli, P.}, {N\o{}rgaard-Nielsen, H. U.}, {Noviello, F.},
  {Novikov, D.}, {Novikov, I.}, {Oxborrow, C. A.}, {Pagano, L.}, {Pajot, F.},
  {Paladini, R.}, {Paoletti, D.}, {Patanchon, G.}, {Pearson, T. J.}, {Peel,
  M.}, {Perdereau, O.}, {Perrotta, F.}, {Piacentini, F.}, {Piat, M.},
  {Pierpaoli, E.}, {Pietrobon, D.}, {Plaszczynski, S.}, {Pointecouteau, E.},
  {Polenta, G.}, {Ponthieu, N.}, {Popa, L.}, {Pratt, G. W.}, {Prunet, S.},
  {Puget, J.-L.}, {Rachen, J. P.}, {Rebolo, R.}, {Reich, W.}, {Reinecke, M.},
  {Remazeilles, M.}, {Renault, C.}, {Ricciardi, S.}, {Riller, T.},
  {Ristorcelli, I.}, {Rocha, G.}, {Rosset, C.}, {Roudier, G.},
  {Rubi\~no-Mart\'{\i}n, J. A.}, {Rusholme, B.}, {Sandri, M.}, {Savini, G.},
  {Scott, D.}, {Spencer, L. D.}, {Stolyarov, V.}, {Sutton, D.}, {Suur-Uski,
  A.-S.}, {Sygnet, J.-F.}, {Tauber, J. A.}, {Tavagnacco, D.}, {Terenzi, L.},
  {Tibbs, C. T.}, {Toffolatti, L.}, {Tomasi, M.}, {Tristram, M.}, {Tucci, M.},
  {Valenziano, L.}, {Valiviita, J.}, {Van Tent, B.}, {Varis, J.}, {Verstraete,
  L.}, {Vielva, P.}, {Villa, F.}, {Wandelt, B. D.}, {Watson, R.}, {Wilkinson,
  A.}, {Ysard, N.}, {Yvon, D.}, {Zacchei, A.}, \& {Zonca,
  A.}}]{Planck-Collaboration2014a}
---. 2014, A\&A, 565, A103

\bibitem[{{Planck Collaboration} {et~al.}(2016{\natexlab{a}}){Planck
  Collaboration}, {Adam, R.}, {Ade, P. A. R.}, {Aghanim, N.}, {Alves, M. I.
  R.}, {Arnaud, M.}, {Ashdown, M.}, {Aumont, J.}, {Baccigalupi, C.}, {Banday,
  A. J.}, {Barreiro, R. B.}, {Bartlett, J. G.}, {Bartolo, N.}, {Battaner, E.},
  {Benabed, K.}, {Beno\^{\i}t, A.}, {Benoit-L\'evy, A.}, {Bernard, J.-P.},
  {Bersanelli, M.}, {Bielewicz, P.}, {Bock, J. J.}, {Bonaldi, A.}, {Bonavera,
  L.}, {Bond, J. R.}, {Borrill, J.}, {Bouchet, F. R.}, {Boulanger, F.},
  {Bucher, M.}, {Burigana, C.}, {Butler, R. C.}, {Calabrese, E.}, {Cardoso,
  J.-F.}, {Catalano, A.}, {Challinor, A.}, {Chamballu, A.}, {Chary, R.-R.},
  {Chiang, H. C.}, {Christensen, P. R.}, {Clements, D. L.}, {Colombi, S.},
  {Colombo, L. P. L.}, {Combet, C.}, {Couchot, F.}, {Coulais, A.}, {Crill, B.
  P.}, {Curto, A.}, {Cuttaia, F.}, {Danese, L.}, {Davies, R. D.}, {Davis, R.
  J.}, {de Bernardis, P.}, {de Rosa, A.}, {de Zotti, G.}, {Delabrouille, J.},
  {D\'esert, F.-X.}, {Dickinson, C.}, {Diego, J. M.}, {Dole, H.}, {Donzelli,
  S.}, {Dor\'e, O.}, {Douspis, M.}, {Ducout, A.}, {Dupac, X.}, {Efstathiou,
  G.}, {Elsner, F.}, {En\ss{}lin, T. A.}, {Eriksen, H. K.}, {Falgarone, E.},
  {Fergusson, J.}, {Finelli, F.}, {Forni, O.}, {Frailis, M.}, {Fraisse, A. A.},
  {Franceschi, E.}, {Frejsel, A.}, {Galeotta, S.}, {Galli, S.}, {Ganga, K.},
  {Ghosh, T.}, {Giard, M.}, {Giraud-H\'eraud, Y.}, {Gjerl\o{}w, E.},
  {Gonz\'alez-Nuevo, J.}, {G\'orski, K. M.}, {Gratton, S.}, {Gregorio, A.},
  {Gruppuso, A.}, {Gudmundsson, J. E.}, {Hansen, F. K.}, {Hanson, D.},
  {Harrison, D. L.}, {Helou, G.}, {Henrot-Versill\'e, S.},
  {Hern\'andez-Monteagudo, C.}, {Herranz, D.}, {Hildebrandt, S. R.}, {Hivon,
  E.}, {Hobson, M.}, {Holmes, W. A.}, {Hornstrup, A.}, {Hovest, W.},
  {Huffenberger, K. M.}, {Hurier, G.}, {Jaffe, A. H.}, {Jaffe, T. R.}, {Jones,
  W. C.}, {Juvela, M.}, {Keih\"anen, E.}, {Keskitalo, R.}, {Kisner, T. S.},
  {Kneissl, R.}, {Knoche, J.}, {Kunz, M.}, {Kurki-Suonio, H.}, {Lagache, G.},
  {L\"ahteenm\"aki, A.}, {Lamarre, J.-M.}, {Lasenby, A.}, {Lattanzi, M.},
  {Lawrence, C. R.}, {Le Jeune, M.}, {Leahy, J. P.}, {Leonardi, R.},
  {Lesgourgues, J.}, {Levrier, F.}, {Liguori, M.}, {Lilje, P. B.},
  {Linden-V\o{}rnle, M.}, {L\'opez-Caniego, M.}, {Lubin, P. M.},
  {Mac\'{\i}as-P\'erez, J. F.}, {Maggio, G.}, {Maino, D.}, {Mandolesi, N.},
  {Mangilli, A.}, {Maris, M.}, {Marshall, D. J.}, {Martin, P. G.},
  {Mart\'{\i}nez-Gonz\'alez, E.}, {Masi, S.}, {Matarrese, S.}, {McGehee, P.},
  {Meinhold, P. R.}, {Melchiorri, A.}, {Mendes, L.}, {Mennella, A.},
  {Migliaccio, M.}, {Mitra, S.}, {Miville-Desch\^enes, M.-A.}, {Moneti, A.},
  {Montier, L.}, {Morgante, G.}, {Mortlock, D.}, {Moss, A.}, {Munshi, D.},
  {Murphy, J. A.}, {Naselsky, P.}, {Nati, F.}, {Natoli, P.}, {Netterfield, C.
  B.}, {N\o{}rgaard-Nielsen, H. U.}, {Noviello, F.}, {Novikov, D.}, {Novikov,
  I.}, {Orlando, E.}, {Oxborrow, C. A.}, {Paci, F.}, {Pagano, L.}, {Pajot, F.},
  {Paladini, R.}, {Paoletti, D.}, {Partridge, B.}, {Pasian, F.}, {Patanchon,
  G.}, {Pearson, T. J.}, {Perdereau, O.}, {Perotto, L.}, {Perrotta, F.},
  {Pettorino, V.}, {Piacentini, F.}, {Piat, M.}, {Pierpaoli, E.}, {Pietrobon,
  D.}, {Plaszczynski, S.}, {Pointecouteau, E.}, {Polenta, G.}, {Pratt, G. W.},
  {Pr\'ezeau, G.}, {Prunet, S.}, {Puget, J.-L.}, {Rachen, J. P.}, {Reach, W.
  T.}, {Rebolo, R.}, {Reinecke, M.}, {Remazeilles, M.}, {Renault, C.}, {Renzi,
  A.}, {Ristorcelli, I.}, {Rocha, G.}, {Rosset, C.}, {Rossetti, M.}, {Roudier,
  G.}, {Rubi\~no-Mart\'{\i}n, J. A.}, {Rusholme, B.}, {Sandri, M.}, {Santos,
  D.}, {Savelainen, M.}, {Savini, G.}, {Scott, D.}, {Seiffert, M. D.},
  {Shellard, E. P. S.}, {Spencer, L. D.}, {Stolyarov, V.}, {Stompor, R.},
  {Strong, A. W.}, {Sudiwala, R.}, {Sunyaev, R.}, {Sutton, D.}, {Suur-Uski,
  A.-S.}, {Sygnet, J.-F.}, {Tauber, J. A.}, {Terenzi, L.}, {Toffolatti, L.},
  {Tomasi, M.}, {Tristram, M.}, {Tucci, M.}, {Tuovinen, J.}, {Umana, G.},
  {Valenziano, L.}, {Valiviita, J.}, {Van Tent, F.}, {Vielva, P.}, {Villa, F.},
  {Wade, L. A.}, {Wandelt, B. D.}, {Wehus, I. K.}, {Wilkinson, A.}, {Yvon, D.},
  {Zacchei, A.}, \& {Zonca, A.}}]{Planck-Collaboration2016d}
{Planck Collaboration}, {Adam, R.}, {Ade, P. A. R.}, {et~al.}
  2016{\natexlab{a}}, A\&A, 594, A10

\bibitem[{{Planck Collaboration} {et~al.}(2016{\natexlab{b}}){Planck
  Collaboration}, {Ade, P. A. R.}, {Aghanim, N.}, {Alves, M. I. R.}, {Arnaud,
  M.}, {Ashdown, M.}, {Aumont, J.}, {Baccigalupi, C.}, {Banday, A. J.},
  {Barreiro, R. B.}, {Bartlett, J. G.}, {Bartolo, N.}, {Battaner, E.},
  {Benabed, K.}, {Beno\^{\i}t, A.}, {Benoit-L\'evy, A.}, {Bernard, J.-P.},
  {Bersanelli, M.}, {Bielewicz, P.}, {Bock, J. J.}, {Bonaldi, A.}, {Bonavera,
  L.}, {Bond, J. R.}, {Borrill, J.}, {Bouchet, F. R.}, {Boulanger, F.},
  {Bucher, M.}, {Burigana, C.}, {Butler, R. C.}, {Calabrese, E.}, {Cardoso,
  J.-F.}, {Catalano, A.}, {Challinor, A.}, {Chamballu, A.}, {Chary, R.-R.},
  {Chiang, H. C.}, {Christensen, P. R.}, {Colombi, S.}, {Colombo, L. P. L.},
  {Combet, C.}, {Couchot, F.}, {Coulais, A.}, {Crill, B. P.}, {Curto, A.},
  {Cuttaia, F.}, {Danese, L.}, {Davies, R. D.}, {Davis, R. J.}, {de Bernardis,
  P.}, {de Rosa, A.}, {de Zotti, G.}, {Delabrouille, J.}, {Delouis, J.-M.},
  {D\'esert, F.-X.}, {Dickinson, C.}, {Diego, J. M.}, {Dole, H.}, {Donzelli,
  S.}, {Dor\'e, O.}, {Douspis, M.}, {Ducout, A.}, {Dupac, X.}, {Efstathiou,
  G.}, {Elsner, F.}, {En\ss{}lin, T. A.}, {Eriksen, H. K.}, {Falgarone, E.},
  {Fergusson, J.}, {Finelli, F.}, {Forni, O.}, {Frailis, M.}, {Fraisse, A. A.},
  {Franceschi, E.}, {Frejsel, A.}, {Galeotta, S.}, {Galli, S.}, {Ganga, K.},
  {Ghosh, T.}, {Giard, M.}, {Giraud-H\'eraud, Y.}, {Gjerl\o{}w, E.},
  {Gonz\'alez-Nuevo, J.}, {G\'orski, K. M.}, {Gratton, S.}, {Gregorio, A.},
  {Gruppuso, A.}, {Gudmundsson, J. E.}, {Hansen, F. K.}, {Hanson, D.},
  {Harrison, D. L.}, {Helou, G.}, {Henrot-Versill\'e, S.},
  {Hern\'andez-Monteagudo, C.}, {Herranz, D.}, {Hildebrandt, S. R.}, {Hivon,
  E.}, {Hobson, M.}, {Holmes, W. A.}, {Hornstrup, A.}, {Hovest, W.},
  {Huffenberger, K. M.}, {Hurier, G.}, {Jaffe, A. H.}, {Jaffe, T. R.}, {Jones,
  W. C.}, {Juvela, M.}, {Keih\"anen, E.}, {Keskitalo, R.}, {Kisner, T. S.},
  {Kneissl, R.}, {Knoche, J.}, {Kunz, M.}, {Kurki-Suonio, H.}, {Lagache, G.},
  {L\"ahteenm\"aki, A.}, {Lamarre, J.-M.}, {Lasenby, A.}, {Lattanzi, M.},
  {Lawrence, C. R.}, {Leahy, J. P.}, {Leonardi, R.}, {Lesgourgues, J.},
  {Levrier, F.}, {Liguori, M.}, {Lilje, P. B.}, {Linden-V\o{}rnle, M.},
  {L\'opez-Caniego, M.}, {Lubin, P. M.}, {Mac\'{\i}as-P\'erez, J. F.}, {Maggio,
  G.}, {Maino, D.}, {Mandolesi, N.}, {Mangilli, A.}, {Maris, M.}, {Marshall, D.
  J.}, {Martin, P. G.}, {Mart\'{\i}nez-Gonz\'alez, E.}, {Masi, S.}, {Matarrese,
  S.}, {McGehee, P.}, {Meinhold, P. R.}, {Melchiorri, A.}, {Mendes, L.},
  {Mennella, A.}, {Migliaccio, M.}, {Mitra, S.}, {Miville-Desch\^enes, M.-A.},
  {Moneti, A.}, {Montier, L.}, {Morgante, G.}, {Mortlock, D.}, {Moss, A.},
  {Munshi, D.}, {Murphy, J. A.}, {Nati, F.}, {Natoli, P.}, {Netterfield, C.
  B.}, {N\o{}rgaard-Nielsen, H. U.}, {Noviello, F.}, {Novikov, D.}, {Novikov,
  I.}, {Orlando, E.}, {Oxborrow, C. A.}, {Paci, F.}, {Pagano, L.}, {Pajot, F.},
  {Paladini, R.}, {Paoletti, D.}, {Partridge, B.}, {Pasian, F.}, {Patanchon,
  G.}, {Pearson, T. J.}, {Peel, M.}, {Perdereau, O.}, {Perotto, L.}, {Perrotta,
  F.}, {Pettorino, V.}, {Piacentini, F.}, {Piat, M.}, {Pierpaoli, E.},
  {Pietrobon, D.}, {Plaszczynski, S.}, {Pointecouteau, E.}, {Polenta, G.},
  {Pratt, G. W.}, {Pr\'ezeau, G.}, {Prunet, S.}, {Puget, J.-L.}, {Rachen, J.
  P.}, {Reach, W. T.}, {Rebolo, R.}, {Reinecke, M.}, {Remazeilles, M.},
  {Renault, C.}, {Renzi, A.}, {Ristorcelli, I.}, {Rocha, G.}, {Rosset, C.},
  {Rossetti, M.}, {Roudier, G.}, {Rubi\~no-Mart\'{\i}n, J. A.}, {Rusholme, B.},
  {Sandri, M.}, {Santos, D.}, {Savelainen, M.}, {Savini, G.}, {Scott, D.},
  {Seiffert, M. D.}, {Shellard, E. P. S.}, {Spencer, L. D.}, {Stolyarov, V.},
  {Stompor, R.}, {Strong, A. W.}, {Sudiwala, R.}, {Sunyaev, R.}, {Sutton, D.},
  {Suur-Uski, A.-S.}, {Sygnet, J.-F.}, {Tauber, J. A.}, {Terenzi, L.},
  {Toffolatti, L.}, {Tomasi, M.}, {Tristram, M.}, {Tucci, M.}, {Tuovinen, J.},
  {Umana, G.}, {Valenziano, L.}, {Valiviita, J.}, {Van Tent, F.}, {Vidal, M.},
  {Vielva, P.}, {Villa, F.}, {Wade, L. A.}, {Wandelt, B. D.}, {Watson, R.},
  {Wehus, I. K.}, {Wilkinson, A.}, {Yvon, D.}, {Zacchei, A.}, \& {Zonca,
  A.}}]{Planck-Collaboration2016e}
{Planck Collaboration}, {Ade, P. A. R.}, {Aghanim, N.}, {et~al.}
  2016{\natexlab{b}}, A\&A, 594, A25

\bibitem[{Poidevin {et~al.}(2023)Poidevin, G\'{e}nova-Santos, Rubi\~{n}o
  Mart\'{i}n, L\'{o}pez-Caraballo, Watson, Artal, Ashdown, Barreiro, Casas,
  de~la Hoz, Fern\'{a}ndez-Torreiro, Guidi, Herranz, Hoyland, Lasenby,
  Martinez-Gonzalez, Peel, Piccirillo, Rebolo, Ruiz-Granados, Tramonte,
  Vansyngel, \& Vielva}]{Poidevin2023}
Poidevin, F., G\'{e}nova-Santos, R.~T., Rubi\~{n}o Mart\'{i}n, J.~A., {et~al.}
  2023, Monthly Notices of the Royal Astronomical Society, 519, 3481

\bibitem[{{Press} {et~al.}(1992){Press}, {Teukolsky}, {Vetterling}, \&
  {Flannery}}]{Press1992}
{Press}, W.~H., {Teukolsky}, S.~A., {Vetterling}, W.~T., \& {Flannery}, B.~P.
  1992, {Numerical recipes in FORTRAN. The art of scientific computing}
  (Cambridge University Press)

\bibitem[{{Privon} {et~al.}(2020){Privon}, {Ricci}, {Aalto}, {Viti}, {Armus},
  {D{\'\i}az-Santos}, {Gonz{\'a}lez-Alfonso}, {Iwasawa}, {Jeff}, {Treister},
  {Bauer}, {Evans}, {Garg}, {Herrero-Illana}, {Mazzarella}, {Larson}, {Blecha},
  {Barcos-Mu{\~n}oz}, {Charmandaris}, {Stierwalt}, \&
  {P{\'e}rez-Torres}}]{Privon2020}
{Privon}, G.~C., {Ricci}, C., {Aalto}, S., {et~al.} 2020, \apj, 893, 149

\bibitem[{Remazeilles {et~al.}(2016)Remazeilles, Dickinson, Eriksen, \&
  Wehus}]{Remazeilles2016}
Remazeilles, M., Dickinson, C., Eriksen, H. K.~K., \& Wehus, I.~K. 2016,
  Monthly Notices of the Royal Astronomical Society, 458, 2032

\bibitem[{{Rubi{\~n}o-Mart{\'\i}n} {et~al.}(2023){Rubi{\~n}o-Mart{\'\i}n},
  {Guidi}, {G{\'e}nova-Santos}, {Harper}, {Herranz}, {Hoyland}, {Lasenby},
  {Poidevin}, {Rebolo}, {Ruiz-Granados}, {Vansyngel}, {Vielva}, {Watson},
  {Artal}, {Ashdown}, {Barreiro}, {Bilbao-Ahedo}, {Casas}, {Casaponsa},
  {Cepeda-Arroita}, {de la Hoz}, {Dickinson}, {Fern{\'a}ndez-Cobos},
  {Fern{\'a}ndez-Torreiro}, {Gonz{\'a}lez-Gonz{\'a}lez},
  {Hern{\'a}ndez-Monteagudo}, {L{\'o}pez-Caniego}, {L{\'o}pez-Caraballo},
  {Mart{\'\i}nez-Gonz{\'a}lez}, {Peel}, {Pel{\'a}ez-Santos}, {Perrott},
  {Piccirillo}, {Razavi-Ghods}, {Scott}, {Titterington}, {Tramonte}, \&
  {Vignaga}}]{Rubino-Martin2023}
{Rubi{\~n}o-Mart{\'\i}n}, J.~A., {Guidi}, F., {G{\'e}nova-Santos}, R.~T.,
  {et~al.} 2023, \mnras, 519, 3383

\bibitem[{Tibbs {et~al.}(2011)Tibbs, Flagey, Paladini, Compi{\~A}{\v s}gne,
  Shenoy, Carey, Noriega-Crespo, Dickinson, Ali-Ha{\~A}¯moud, Casassus,
  Cleary, Davies, Davis, Hirata, \& Watson}]{Tibbs2011}
Tibbs, C.~T., Flagey, N., Paladini, R., {et~al.} 2011, Monthly Notices of the
  Royal Astronomical Society, 418, 1889

\bibitem[{{Tielens}(2008)}]{Tielens2008}
{Tielens}, A.~G.~G.~M. 2008, \araa, 46, 289

\bibitem[{{Weingartner} \& {Draine}(2001)}]{Weingartner2001}
{Weingartner}, J.~C., \& {Draine}, B.~T. 2001, \apjs, 134, 263

\end{thebibliography}
\bibliographystyle{aasjournal}

\end{document}